# Compact Orthomode Transducers Using Digital Polarization Synthesis

Matthew A. Morgan - matt.morgan@nrao.edu
J. Richard Fisher - rfisher@nrao.edu
Tod A. Boyd - tboyd@nrao.edu

National Radio Astronomy Observatory
1180 Boxwood Estate Rd.
Charlottesville, VA. 22903

**Abstract**

In this paper we present a novel class of compact orthomode transducers which use digital calibration to synthesize the desired polarization vectors while maintaining high isolation and minimizing mass and volume. These digital orthomode transducers consist of an arbitrary number of planar probes in a circular waveguide, each of which is connected to an independent receiver chain designed for stability of complex gain. The outputs of each receiver chain are then digitized and combined numerically with calibrated, complex coefficients. Measurements on two prototype digital orthomode transducers, one with three probes and one with four, show better than 50 dB polarization isolation over a 10 C temperature range with a single calibration.

keywords: orthomode transducer (OMT), polarimetry, polarization

## I. Introduction

One of the most difficult elements to integrate within a compact, multi-beam, radio astronomy receiver is the orthomode transducer. Often bulky and geometrically awkward, it tends to constrain fabrication in ways that are not compatible with efficient integration of other components, such as the low-noise amplifiers (LNAs). In this paper we develop a modification of the planar orthomode transducer geometry which permits easy integration of the cold electronics in a package not much larger than the flange of the circular waveguide which serves as its input.

This is the second phase in a long-term program to re-optimize high-performance radio astronomy receiver architecture by leveraging the advances in modern digital computing [1]. Intrinsic to that program are the ideas that optimal performance is achieved if the signal is digitized as close to the antenna feed as possible, and that this inevitably involves transferring some functionality from the analog into the digital domain. In the first part of the program, a sideband-separating downconverter was developed in which the final sideband separation was performed digitally using calibrated, complex weighting coefficients to achieve an unprecedented level of performance with exceptional stability [2]. With this work we build upon that result by adding calibrated polarization isolation to the digital processor.

## II. Principles of Operation

Photographs of two prototype digital orthomode transducers are shown in Fig. 1. They comprise three and four planar probes, respectively, extending into a 0.92-inch diameter circular waveguide with a flat-bottom backshort. For this experiment, the digital orthomode transducers were designed to be bolted directly onto the output of a corrugated circular feedhorn operating in X-Band (8-12 GHz).

The four-probe version resembles, from the outside at least, the planar orthomode transducers under development concurrently by a number of different research groups around the world [3]-[5]. The details of the interior, however, are quite different. As is evident in Fig. 1d, there are no analog baluns to recombine the signals from opposing probes – baluns, which must inevitably increase the receiver noise temperature through insertion loss and which over a broad bandwidth have amplitude and phase imbalances that contribute to gain ripple and possibly polarization coupling. Instead, a cascaded pair of integrated, low-noise, MMIC amplifiers is connected directly to the terminal of each planar probe. Each of these small modules thus represents the entire cooled electronics package of an X-Band receiver (though cooling was not performed for this early experiment). Each LNA drives an independent receiver chain, specially designed for stability with downconverters based on the previous work [2]. The outputs of all receiver chains are then digitized and processed numerically to reconstruct the two polarizations with extraordinary precision.

The first question we must address is whether one should even expect the three-probe arrangement to work at all, as its operation is far less intuitive than the four-probe counterpart. In fact, we will show that a digital orthomode transducer constructed with $N$ probes distributed uniformly around a circle must work for $N$ greater than 2 over a standard waveguide bandwidth, and may work for $N$ equal to 2 under specific circumstances.

To first order, each probe couples to a vector component of the E-field that is aligned parallel to its axis. Therefore, any two probes that are not degenerate (i.e., not parallel to one another) provide a vector-basis in terms of which any incoming E-field polarization can be expressed. The decomposition of the E-field into these component vectors is most intuitive if the probes are orthogonal, but this is by no means a requirement.

Additionally, we would prefer that any incident wave, no matter what its polarization, does not reflect back out of the structure in a different mode. This would introduce a loss mechanism which could increase the noise temperature of the receiver, and interactions between the digital orthomode transducer and the feedhorn or the external antenna structure in this mode may create polarization-mixing resonances on a length-scale that cannot be counted on to be stable. In contrast, reflections which are not subject to mode-conversions at either end can in theory be matched, and any standing waves contribute only to gain ripple, not cross-polar leakage.

Fortunately, the symmetry of the structure ensures that no such mode-conversion will take place. Consider the *x* and *y* polarizations impinging on the three-probe digital orthomode transducer shown diagrammatically in Fig. 2a. For simplicity, we assume that one of the probes lies directly on the *x* axis, and that all probes are terminated with equal impedances. By symmetry, it is clear that no incoming wave polarized in the *x* direction may couple into the *y* polarization. By reciprocity then, no incoming wave polarized in the *y* direction may couple into the *x* polarization. Since all linear polarizations may be broken down into components of *x* and *y*, and the *x* and *y* components do not couple into one another, it follows that no linear polarization, no matter what its orientation with respect to the probes, may convert into another linear polarization by its interaction with them. This argument holds for any arrangement of probes (not necessarily uniform in distribution) with at least one symmetry axis. We note that even a two-probe arrangement, Fig. 2b, satisfies this criterion.

What about higher-order modes? As shown in Fig. 3, the next mode in circular waveguide ($TM_{01}$) after the two linear modes (both $TE_{11}$) has rotational-symmetry, meaning that it is unchanged by arbitrary rotation about the longitudinal axis of the waveguide. If we assume that this mode impinging onto the three-probe structure does couple into a linear mode, then it follows by symmetry that it must also couple to an identical linear mode rotated by 120°, and to yet a third linear mode 120° beyond that, all with equal amplitude and phase. The superposition of these three modes sum to zero, however, disproving the assumption. Thus, the rotationally-symmetric $TM_{01}$ mode may not couple into any linear mode, and by reciprocity no linear mode may couple into the rotationally symmetric mode, provided that the probes are arranged at uniform intervals around the circle. Note that the two-probe case, where the probes are separated by 90°, does not satisfy this criterion, and thus is subject to $TM_{01}$ mode conversion unless the waveguide geometry puts this mode below cutoff (as is true, for example, in a standard quad-ridge orthomode transducer [6] which is topologically similar to the two-probe case).

Finally, one might ask whether incident waves of every possible polarization may all be simultaneously impedance matched given equal terminations on all probes. Once again, symmetry guarantees that this is the case. Using the three-probe arrangement once more as an example, we assume that the structure has been matched to at least one linear polarization using equal terminations on all probes. It follows that the structure must also be matched to a linear polarization rotated 120° from the first. Since any other linear polarization is but a weighted sum of these two – they may not be orthogonal, but they still span the vector space – any linear polarization whatsoever is also matched by this structure. Thus we can be confident that if any one polarization is matched, then all polarizations are matched. This holds for any uniform distribution of probes of *N* greater than 2, but also for the two-probe case shown in Fig. 2b, since the matching problem for a linear polarization aligned parallel to one probe is but a mirror image of the same problem for matching in a polarization aligned with the other.

The simulated return loss at the waveguide flange for the three- and four-probe configurations is shown in Fig. 4. The three-probe version was optimized for 30 dB return loss over a relatively narrow band, whereas the four-probe version was optimized for 20 dB over the full 8-12 GHz bandwidth. These simulations assume that the terminating impedance on each probe is 50 Ω. In practice, the input match will be dominated by that of the LNAs.

## III. Analysis

The input-output behavior of a digital orthomode transducer consisting of *N* probes, whatever their geometry and however they are distributed, may be described by the following equation.

$$\begin{bmatrix} v_1 \\ v_2 \\ \vdots \\ v_N \end{bmatrix} = \begin{bmatrix} a_{1x}e^{j\theta_{1x}} & a_{1y}e^{j\theta_{1y}} \\ a_{2x}e^{j\theta_{2x}} & a_{2y}e^{j\theta_{2y}} \\ \vdots & \vdots \\ a_{Nx}e^{j\theta_{Nx}} & a_{Ny}e^{j\theta_{Ny}} \end{bmatrix} \begin{bmatrix} s_x \\ s_y \end{bmatrix} \tag{1}$$

or

$$V = GS \tag{2}$$

where *V* is the output voltage vector from all *N* receiver channels connected to the digital orthomode transducer, *S* is the input signal vector in x and y polarizations, and *G* is the complex gain matrix that defines the relationship between them. The amplitude, $a_{ix}$, represents the net effective amplitude response of probe *i* to *x*-polarized input signals, along with the gains of all the electronics in the receiver chain attached to it. Likewise, $\theta_{ix}$ is the net effective phase response, and so on for the remaining *x* and *y* terms. The only assumption in this formulation is that the digital orthomode transducer is linear.

The input signal may be broken down as

$$S = s_0 \begin{bmatrix} \cos\alpha \\ \sin\alpha \end{bmatrix} e^{j\beta} \tag{3}$$

where $s_0$ is the signal amplitude, $\alpha$ is the linear polarization angle with respect to the $x$ axis, and $\beta$ is an arbitrary phase constant.

For astronomical observation, one would like to reverse the relationship in (2) – that is, given a measured output vector $V$, we would like to calculate the input signal vector, $S$, which produced it. Direct inversion is not possible, as the gain matrix is not generally square (unless $N$=2). In mathematical terms, the system is over-constrained. Only two probes are required to determine everything needed to be known about the input signal, but we have more than that. The excess information could possibly be ignored, but a more elegant solution is to use the Moore-Penrose Pseudo-Inverse [7], given by

$$H = \left(G^T G\right)^{-1} G^T . \tag{4}$$

This provides for a least-squares fit of the solution, $S'$, to the data, $V$,

$$S' = HV . \tag{5}$$

The reader may verify using the expression for $H$ in (4) that the product $HG$ is equal to the identity matrix. In practice, the pseudo-inverse, $H$, is frequency dependant, and estimation of the sky-signal vector from phasor voltages may take place anywhere in the digital processing system where frequency spectra are available, such as in the spectrometer or correlator after an FFT. Often, only cross-product spectra are available, in which case we define a measurement matrix as

$$M = \left\langle VV^* \right\rangle \tag{6}$$

where the angle brackets denote the time-average. This is a useful technique as the undetermined phase constant, $\beta$, drops out. We may further observe that

$$\left\langle S'S'^* \right\rangle = \begin{bmatrix} |s'_x|^2 & s'_x s'^*_y \\ s'_y s'^*_x & |s'_y|^2 \end{bmatrix} = \left\langle HV(HV)^* \right\rangle = HMH^* . \tag{7}$$

Therefore, the power levels of the two orthogonal components of the input polarization are given by the diagonal elements of the matrix $HMH^*$.

Linear polarizations may be converted to circular polarizations by observing that

$$\begin{bmatrix} s_l \\ s_r \end{bmatrix} = \begin{bmatrix} 1 & -j \\ 1 & +j \end{bmatrix} = \begin{bmatrix} s_x \\ s_y \end{bmatrix} \tag{8a}$$

$$\begin{bmatrix} s_x \\ s_y \end{bmatrix} = \frac{1}{2} \begin{bmatrix} 1 & 1 \\ j & -j \end{bmatrix} \begin{bmatrix} s_l \\ s_r \end{bmatrix} \tag{8b}$$

and solving for a new pseudo-inverse matrix

$$H_{circular} = \begin{bmatrix} 1 & -j \\ 1 & +j \end{bmatrix} H_{linear} . \tag{9}$$

We may also rotate the linear polarization to any orientation required,

$$H_\gamma = \begin{bmatrix} \cos\gamma & \sin\gamma \\ -\sin\gamma & \cos\gamma \end{bmatrix} H . \tag{10}$$

The ideal gain matrices for the three- and four-probe digital orthomode transducers shown in Fig. 1, normalized to unity gain and neglecting differential contributions from the receiver components behind the probes, may be derived by projection of the linear polarization vector onto the axis of each probe,

$$G_3 = \begin{bmatrix} 1 & 0 \\ -\frac{1}{2} & \frac{\sqrt{3}}{2} \\ -\frac{1}{2} & -\frac{\sqrt{3}}{2} \end{bmatrix} \tag{11a}$$

$$G_4 = \begin{bmatrix} 1 & 0 \\ 0 & 1 \\ -1 & 0 \\ 0 & -1 \end{bmatrix} \tag{11b}$$

where it has been assumed that probe 1 is in line with the $x$ axis. The pseudo-inverse, $H$, may be calculated using (4),

$$H_3 = \begin{bmatrix} \frac{2}{3} & -\frac{1}{3} & -\frac{1}{3} \\ 0 & \frac{\sqrt{3}}{3} & -\frac{\sqrt{3}}{3} \end{bmatrix} \tag{12a}$$

$$H_4 = \begin{bmatrix} \frac{1}{2} & 0 & -\frac{1}{2} & 0 \\ 0 & \frac{1}{2} & 0 & -\frac{1}{2} \end{bmatrix} . \tag{12b}$$

More generally, for an $N$-probe digital orthomode transducer, where the distribution of the probes is uniform and probe 1 is rotated by an angle $\psi$ from the $x$ axis,

$$G_{N,\psi} = \begin{bmatrix} \cos\left(0 \cdot \frac{2\pi}{N} + \psi\right) & \sin\left(0 \cdot \frac{2\pi}{N} + \psi\right) \\ \cos\left(1 \cdot \frac{2\pi}{N} + \psi\right) & \sin\left(1 \cdot \frac{2\pi}{N} + \psi\right) \\ \vdots & \vdots \\ \cos\left(N \cdot \frac{2\pi}{N} + \psi\right) & \sin\left(N \cdot \frac{2\pi}{N} + \psi\right) \end{bmatrix} \tag{13a}$$

$$H_{N,\psi} = \frac{2}{N} \begin{bmatrix} \cos\left(0 \cdot \frac{2\pi}{N} + \psi\right) & \cdots & \cos\left(N \cdot \frac{2\pi}{N} + \psi\right) \\ \sin\left(0 \cdot \frac{2\pi}{N} + \psi\right) & \cdots & \sin\left(N \cdot \frac{2\pi}{N} + \psi\right) \end{bmatrix} . \tag{13b}$$

## IV. Numerical Calibration

To obtain precise reconstruction of the polarization vector, $S$, we must first determine the elements of the gain matrix, $G$, by calibration measurements. Although, in principle, calibration may be performed in either linear or circular polarization, it is usually more straightforward to generate linear polarization in the lab, using a waveguide taper or polarizing grid.

Initially, we take two measurements, first with an input test signal aligned with the $x$ axis ($\alpha$=0), and second with an input test signal aligned with the $y$ axis ($\alpha$=π/2), and examine the cross-product matrix, $M$, defined by (6). The elements of this matrix, $m_{ik}$, are the measured cross products of the $i^{\text{th}}$ and $k^{\text{th}}$ output voltages, and are given by

$$m_{ik}(\alpha) = \left\langle v_i(\alpha) v_k^*(\alpha) \right\rangle \tag{14a}$$

$$= s_0^2 \left( a_{ix} e^{j\theta_{ix}} \cos\alpha + a_{iy} e^{j\theta_{iy}} \sin\alpha \right) \left( a_{kx} e^{j\theta_{kx}} \cos\alpha + a_{ky} e^{j\theta_{ky}} \sin\alpha \right). \tag{14b}$$

Using this expression, we can derive the gain terms from the measurements as follows,

$$a_{ix} = s_0^{-1} \sqrt{m_{ii}(0)} \tag{15a}$$

$$a_{iy} = s_0^{-1} \sqrt{m_{ii}\left(\tfrac{\pi}{2}\right)} \tag{15b}$$

$$\theta_{ix} - \theta_{kx} = \arctan\left( \frac{\operatorname{Im}\{m_{ik}(0)\}}{\operatorname{Re}\{m_{ik}(0)\}} \right) \tag{15c}$$

$$\theta_{iy} - \theta_{ky} = \arctan\left( \frac{\operatorname{Im}\{m_{ik}\left(\tfrac{\pi}{2}\right)\}}{\operatorname{Re}\{m_{ik}\left(\tfrac{\pi}{2}\right)\}} \right). \tag{15d}$$

Thus, we obtain the magnitudes and relative phases of all terms within each column of the gain matrix (but not the relative phase between the two columns). For each column, a single probe is chosen as the phase reference, so that its phase is zero and the phases of all other terms in the column are set relative to it.

Note that in some cases, during the evaluation of (15c) and (15d), one or both of the probes, $i$ or $k$, may be orthogonal to the input signal. In these situations, the cross-product, $m_{ik}$, will be close to zero, and the extracted relative phase will be noisy (mathematically speaking, the angle is indeterminate). For this reason, it is important to choose a probe for the phase reference in each column that is not orthogonal to the input. Then the remaining probes which are not orthogonal should have well-defined phases relative to it. Only those probes which are orthogonal to the input polarization will have indeterminate phase relative to the reference. This is acceptable since by definition these probes provide no useful information about the given component, $x$ or $y$, of the input signal.

In practice, it is a simple matter for the calibration algorithm to choose the one probe in each column which has the largest gain magnitude, $a_{ix}$ or $a_{iy}$, and thus the strongest correlation to the input signal to use as the reference. The software needs only to do its best to determine the phase angles of the rest of the terms in that column. The ones that are approximately orthogonal could potentially be very noisy, but this should not matter because the entire term will receive a very small weight in the final solution.

Note that it is also possible for a given set of cross-products to be internally inconsistent. This is because there is more information available in the cross-products than is needed to fill in all the unknowns of the linear gain matrix. If there is some systematic error in the determination of the cross-products (such as an offset in the samplers), then the same gain terms derived from different cross-products may not always agree. From (14) we know that the following equalities should hold for a consistent data set

$$\left| m_{ik}(\alpha) \right|^2 = \left| m_{ii}(\alpha) \right| \left| m_{kk}(\alpha) \right| \tag{16a}$$

$$\angle m_{ik}(\alpha) = \angle m_{ij}(\alpha) + \angle m_{jk}(\alpha). \tag{16b}$$

Examining these relationships for consistency can be a useful diagnostic check.

The rough calibration derived thus far is sufficient for most system diagnostics and even many astronomical measurements. Two minor issues remain, however. First, the relative phase between the $x$ and $y$ columns has not been determined. This is usually only important if the user is interested in circular polarization.

The second issue has to do with the orthogonality of the $x$ and $y$ calibrators. Although the interface between the digital orthomode transducer under test and the signal source may be pinned for mechanical accuracy, some residual error due to manufacturing tolerances will always remain. In practice, the axial alignment is usually better than a fraction of degree, but even a half degree misalignment could introduce cross-polar leakage on the order of -40 dB. Since immunity to manufacturing tolerance is one of the chief benefits of a calibrated approach, it is desirable to come up with a way to correct for this. The solution lies in recognizing that if the $x$ and $y$ axes are not orthogonal, then any attempt to synthesize circular polarization will result in measurable ellipticity.

Returning to (2), we know that

$$V = GS = G'S'. \tag{17}$$

In other words, the output voltage vector must be the same whether we pair the real input signal vector, $S$, with the correct gain matrix, $G$, or the estimated signal vector, $S'$, with the partially calibrated gain matrix, $G'$. Allowing for an $x$ versus $y$ phase differential of $\phi$ and a misalignment of the $y$ axis of $\varepsilon$, we can write

$$G'S' = G'\Phi \mathrm{E} S = G' \begin{bmatrix} 1 & 0 \\ 0 & e^{j\phi} \end{bmatrix} \begin{bmatrix} 1 & \sin\varepsilon \\ 0 & \cos\varepsilon \end{bmatrix} \begin{bmatrix} \cos\alpha \\ \sin\alpha \end{bmatrix} \tag{18a}$$

$$\therefore S' = \Phi \mathrm{E} S = \begin{bmatrix} \cos\alpha + \sin\varepsilon \sin\alpha \\ e^{j\phi} \cos\varepsilon \sin\alpha \end{bmatrix}. \tag{18b}$$

Using (8a) we determine the estimated circular components as

$$\begin{bmatrix} s_l' \\ s_r' \end{bmatrix} = \begin{bmatrix} 1 & -j \\ 1 & +j \end{bmatrix} \begin{bmatrix} \cos\alpha + \sin\varepsilon \sin\alpha \\ e^{j\phi} \cos\varepsilon \sin\alpha \end{bmatrix} \tag{19a}$$

$$= \begin{bmatrix} \cos\alpha + \sin\alpha \left( \sin\varepsilon + \sin\phi \cos\varepsilon \right) - j \sin\alpha \cos\phi \cos\varepsilon \\ \cos\alpha + \sin\alpha \left( \sin\varepsilon - \sin\phi \cos\varepsilon \right) + j \sin\alpha \cos\phi \cos\varepsilon \end{bmatrix} \tag{19b}$$

whose magnitudes are

$$\begin{bmatrix} |s_l'|^2 \\ |s_r'|^2 \end{bmatrix} = \begin{bmatrix} 1+\sin(2\varepsilon)\sin^2\alpha\sin\phi+\sin(2\alpha)(\sin\varepsilon+\sin\phi\cos\varepsilon) \\ 1-\sin(2\varepsilon)\sin^2\alpha\sin\phi+\sin(2\alpha)(\sin\varepsilon-\sin\phi\cos\varepsilon) \end{bmatrix}. \qquad (20)$$

The effects of both the orthogonality and phase differentials are most prominent at $\alpha=\pi/4$, halfway between the $x$ and $y$ axes,

$$\begin{bmatrix} |s_l'(\tfrac{\pi}{4})|^2 \\ |s_r'(\tfrac{\pi}{4})|^2 \end{bmatrix} = \begin{bmatrix} 1+\tfrac{1}{2}\sin(2\varepsilon)\sin\phi+\sin\varepsilon+\sin\phi\cos\varepsilon \\ 1-\tfrac{1}{2}\sin(2\varepsilon)\sin\phi+\sin\varepsilon-\sin\phi\cos\varepsilon \end{bmatrix} \qquad (21a)$$

$$|s_l'(\tfrac{\pi}{4})|^2 + |s_r'(\tfrac{\pi}{4})|^2 = 2(1+\sin\varepsilon) \qquad (21b)$$

$$|s_l'(\tfrac{\pi}{4})|^2 - |s_r'(\tfrac{\pi}{4})|^2 = \sin\phi[\sin(2\varepsilon)+2\cos\varepsilon]. \qquad (21c)$$

Finally, we can solve for the orthogonality and phase terms as

$$\sin\varepsilon = \tfrac{1}{2}\left(|s_l'(\tfrac{\pi}{4})|^2 + |s_r'(\tfrac{\pi}{4})|^2\right) - 1 \qquad (22a)$$

$$\sin\phi = \frac{|s_l'(\tfrac{\pi}{4})|^2 - |s_r'(\tfrac{\pi}{4})|^2}{\sin(2\varepsilon)+2\cos\varepsilon}. \qquad (22b)$$

Note that in both cases only the sine of the angle is determined. In the case of the orthogonality term, $\varepsilon$, the value is expected to be very small, so the principle value of the arcsine is the correct one.

On the other hand, the $x$ versus $y$ phase differential, $\phi$, can have vastly different values depending on the choice of reference probes for the $x$ and $y$ columns. To resolve this ambiguity, we must use the cross-product term

$$s_r'(\tfrac{\pi}{4})s_r'^*(\tfrac{\pi}{4}) = \sin^2\varepsilon + \sin\varepsilon + j\cos\phi\left(\cos\varepsilon + \tfrac{1}{2}\sin(2\varepsilon)\right) \qquad (23a)$$

$$\therefore \cos\phi = \frac{2\,\mathrm{Im}\{s_r'(\tfrac{\pi}{4})s_l'^*(\tfrac{\pi}{4})\}}{\sin(2\varepsilon)+2\cos\varepsilon}. \qquad (23b)$$

Having now calculated these terms using a third calibration measurement at $\alpha=\pi/4$, we may apply them to gain matrix by multiplication,

$$G = G'\Phi\mathrm{E} = G'\begin{bmatrix} 1 & 0 \\ 0 & e^{j\phi} \end{bmatrix}\begin{bmatrix} 1 & \sin\varepsilon \\ 0 & \cos\varepsilon \end{bmatrix}. \qquad (24)$$

## V. Error Analysis

To examine the impact of small errors during the extraction of the gain matrix, we may write the sky signals inferred from an observation using an imperfect calibration as

$$S' = HV = \left(G^T G\right)^{-1} G^T G_0 S = (I+E)S \qquad (25)$$

where $G$ is the gain matrix as derived above, which may be in error, $G_0$ is the actual gain matrix of the analog hardware, and $S$ represents the actual input signal from the sky. The error matrix, $E$, is given by

$$E = \left(G^T G\right)^{-1} G^T G_0 - I = \begin{bmatrix} e_{xx} & e_{xy} \\ e_{yx} & e_{yy} \end{bmatrix}. \qquad (26)$$

Note that the main diagonal elements, $e_{xx}$ and $e_{yy}$, represent co-polarization gain errors only, which are automatically corrected by routine astronomical calibration of receiver gain and noise temperature. The counter-diagonal elements, $e_{xy}$ and $e_{yx}$, lead to cross-polarization errors, which are the primary concern here. To evaluate $E$, it is convenient to write the gain matrix as

$$G = G_0 + dG = \lfloor g_x \quad g_y \rfloor. \qquad (27)$$

In this equation, $dG$ is the small error in the derived gain matrix, and $g_x$ and $g_y$ are the two column vectors associated with it – their dimension depending on the number of probes. Substituting this back into (26),

$$E = \left(G^T G\right)^{-1} G^T (G - dG) - I \qquad (28a)$$

$$= -\left(G^T G\right)^{-1} G^T dG \qquad (28b)$$

$$= -\begin{bmatrix} g_x^T g_x & g_x^T g_y \\ g_y^T g_x & g_y^T g_y \end{bmatrix}^{-1} \begin{bmatrix} g_x^T dg_x & g_x^T dg_y \\ g_y^T dg_x & g_y^T dg_y \end{bmatrix}. \qquad (28c)$$

Since $g_x$ and $g_y$ are column vectors, their inner products are equal. In addition, for a well-designed digital orthomode transducer, they are approximately orthogonal and equal in magnitude. That is,

$$g_x^T g_y = g_y^T g_x \approx 0 \qquad (29a)$$

$$g_x^T g_x \approx g_y^T g_y = a^2\,\tfrac{N}{2} \qquad (29b)$$

where $a$ is the nominal voltage gain of a typical receiver channel behind each probe, and $N$ is the number of probes. The reader may verify that these relations hold for all the gain matrices shown in (11) through (13). These properties allow us to simplify the result in (28c),

$$\begin{bmatrix} e_{xx} & e_{xy} \\ e_{yx} & e_{yy} \end{bmatrix} \approx -\left(a^2\,\tfrac{N}{2}\right)^{-1} \begin{bmatrix} g_x^T dg_x & g_x^T dg_y \\ g_y^T dg_x & g_y^T dg_y \end{bmatrix}. \qquad (30)$$

To make use of (30), we must first make assumptions about how a particular mechanism introduces systematic errors into the calibration. For example, are all probes affected by the same error in the same way? Do the errors correlate down a column (on all probes for one polarization) or across a row (on both polarizations for a single probe)? The impact that a particular error will have in the final observation will depend on the answers to these questions. Random, uncorrelated errors, such as might result from simple noisy measurements, may be estimated to add in the root-sum-squares sense,

$$e_{xy} \approx -\left(a^2 \tfrac{N}{2}\right)^{-1}\left(g_x^T dg_y\right) = \left(a^2 \tfrac{N}{2}\right)^{-1}\left(a^2 \tfrac{\sqrt{N}}{2}\zeta\right) = \frac{\zeta}{\sqrt{N}} \cdot \quad (31)$$

where $\zeta$ is the magnitude of the error relative to the average amplitude of the terms in the gain matrix. This shows that isolation in the presence of random noise is limited by

$$Iso_{dB} = 20\log\left|\frac{\zeta}{\sqrt{N}}\right| \cdot \quad (32)$$

where $\zeta$ may be used directly for magnitude errors, or in the case of phase errors,

$$\zeta = \sin\phi_{err} \cdot \quad (33)$$

Thus, random voltage amplitude errors of 1% may limit the isolation of a four probe digital orthomode transducer to -46 dB, while phase errors of half a degree may impose a limit of -47 dB. The number of probes has a very small effect in error sensitivity which scales as the square root of *N*. The difference between three probes and four probes is thus only 1.25 dB. This small advantage of having additional probes would not, in general, offset the increased overhead of processing the additional channels.

## VI. Measurements

In order to test the two prototype digital orthomode transducers, a complete X-Band receiver was constructed, with net RF plus IF gain of 80 dB. Following the digital orthomode transducers themselves is a compact, four-channel downconverter unit, shown in Fig. 5, built using the stable design techniques described in [2]. The measured conversion gain of the downconverter unit as a function of RF and IF is shown in Figs. 6 and 7, respectively.

The completed test receiver, shown in Fig. 8, was small enough to fit on a laboratory hot-plate, allowing the temperature of the unit to be controlled for investigation of calibration stability. The data were collected by an eight-channel analog-to-digital converter, National Instruments model number PXIe-5105, with a sample rate of 60 MS/s. The data were buffered in 0.0.27-second bursts, saved to disk, and later post-processed in software.

RF input at -90 dBm was provided by an Agilent 83640A Synthesizer, set to -60 dBm output with a 30 dB pad. It was injected into the digital orthomode transducer through a transition from coax to WR-90 waveguide. Following that, a stepped taper from rectangular to circular waveguide provided a linear polarization, while an aluminum circular waveguide adapter allowed the interface at the flange to be rotated full-circle in 22.5° increments. The LO was provided to the downconverter by an Agilent E8257D Synthesizer set to +13 dBm output power. No correction for cable losses was made.

The gain matrices resulting from the three measurement calibration procedure described in Section IV, are shown in Figs. 9 through 12. Several things may be noted from these plots. First, the approximate magnitude of the gain terms are in line with the expected values given by (11), where some deviation is expected due to different amplifier gains in the analog hardware after the probes. Second, the terms for which one of the probes is orthogonal to the calibration signal show an intrinsic isolation due to symmetry and manufacturing tolerance on the order of 30 dB.

Additionally, for the three-probe digital orthomode transducer, one of the four channels of the downconverter was left unconnected, revealing that isolation inside the downconverter is better than 70 dB. This highlights an important fact regarding the behavior of the digital orthomode transducers. The same code was used to analyze the data from both modules, with no special provision made for the number of probes or their orientation. It will be shown that this has no obvious effect on the performance of the three-probe digital orthomode transducer. The 'dead channel' simply receives zero (or near-zero) weight in the calibrated solution. This shows that the algorithm is robust to extreme variations in the digital orthomode transducer geometry.

The odd-symmetric nature of the phase curves and the discontinuity at IF=0 is due to the relatively large phase mismatch between channels in the RF circuit, most significantly coming from the coaxial cables between the digital orthomode transducer and the downconverter. RF phase and magnitude differentials have odd symmetry about IF=0, while IF phase and magnitude differentials have even symmetry.

The accuracy of the calibration measurements may be increased by raising the level of the CW test signal (as long as it is somewhat lower than the total system noise power), by increasing the frequency resolution with a longer FFT, or by averaging spectra from successive FFTs on the sample time series. The gain matrices shown in Figs. 9 through 12 were derived by averaging approximately 1000 half-window-overlapped, 32768-sample FFTs (16384 spectral channels) for a total integration time of 0.27 seconds. To achieve the best results, a complex vector sum was performed over a small number of adjacent FFT bins, ensuring that all the power from the test signal was included in the measurement. This was sufficient to achieve isolation on the order of 50 dB.

An important feature of the analog design is that the amplitude and phase curves are smooth throughout their passbands, permitting accurate interpolation of the gain matrix between calibration points and minimizing the total number of complex coefficients that must be stored. The level of interpolation error that can be tolerated for a given isolation requirement may be estimated using (32) and (33). Stability of the amplitude and phase is also key, as it was for the DSSM [2].

After the initial calibration, measurements were taken with the input transition rotated through a full 180° in 22.5° steps. The output voltages were processed to synthesize both linear and circular polarizations in software. A plot of the reconstructed signal strength (averaged across IF frequency) as a function of polarization angle for *x*-linear, *y*-linear, left-circular, and right-circular polarization, along with theoretical predictions, is shown in Fig. 13.

Polarization isolation was measured as the ratio of the two synthesized *x* and *y* output polarizations at the point where one of them should theoretically become null (that is, at 0° and 90° in Fig. 13). The results are plotted versus IF frequency for the two digital orthomode transducers in Figs. 14 and 15. The measurement was first performed immediately after calibration at a temperature of 30 C. This data is shown in the plots with closed markers. The measurement was then repeated after using the hot plate to increase the base temperature of the entire receiver – including all 80 dB of gain, downconversion, and filtering – by 10 C, without recalibrating. This data is shown with open markers. The plot shows that linear polarization isolation is better than 50 dB in both digital orthomode transducers, with almost no measurable degradation due to a temperature change which far exceeds the temperature range that a stabilized receiver would normally experience in practice. It is not known at this time why the three-probe digital orthomode transducer showed a slight degradation, on the order of a few dB, while the four-probe digital orthomode transducer showed none at all.

Axial ratio was measured as the difference between the maximum and minimum synthesized circular polarization amplitude at each IF frequency while rotating the input transition through a 180° turn in 22.5° steps. The results are plotted versus IF in Figs. 16 and 17. As before, the measurements were taken twice, at the calibration temperature of 30 C and again at 40 C without recalibration. In this case, the three-probe and four-probe digital orthomode transducers behaved about the same. The initial axial ratio was better than 0.05 dB after calibration, and remained better than 0.2 dB after the 10 C temperature change.

To ensure that the level of performance achieved with these digital orthomode transducers was not critically dependent on the resolution of the ADC, a subset of the data was reprocessed after masking off different numbers of low order bits, simulating the result that would be obtained with low bit-resolution ADCs. As shown in Figs. 18 and 19, the isolation in the passband remains almost constant even as the number of active bits is reduced to 4. At this level, the analog signal power was such that the ADC behaved like a 1 bit, 2 level sampler (in other words, the amplitude was insufficient to trigger the three most significant bits). The loss of isolation at the band edges is attributable to quantization error distortions from the analog-to-digital conversion process rather than intrinsic isolation properties. The calibration showed that the analog gain at the band edges was approximately 45 dB lower than at the band center due to cutoff of the anti-aliasing filters.

## VII. Uncorrected Errors

It would be dishonest to claim stable polarization isolation on the order of 50 dB, or axial ratio better than 0.05 dB, without at least acknowledging some of the known limitations on the accuracy of the measurements. As described in Section IV, the calibration procedure guarantees that the final *x* and *y* axes are orthogonal, even if the calibration sources themselves were not. It does not, however, guarantee that the two calibration sources are purely linear. They may have a very slight elliptical component to them. In this sense, the data is only as good as the calibration signals provided. Fortunately, we have good reason to expect the calibration signals used in these measurements to be very nearly linear, as this depends only on the symmetry of the WR-90 rectangular waveguide to circular waveguide taper, subject to the manufacturing tolerances involved. Nonetheless, the exact precision of this piece is not known.

Additionally, the orientation of the calibrated *x* and *y* axes with the desired coordinate system of the observation depends on the accuracy of mechanical interfaces, which themselves are probably not as good as that implied by the 50 dB isolation result. The calibration procedure described above fixes the polarization coordinate system to the *x* axis calibrator, and then forces the *y* axis to be perpendicular to that. If a particular angular offset is known or can somehow be determined, a numerical correction may be made quite easily by application of (10).

Finally, in a real-world radio astronomy scenario, the optics associated with the feedhorn, the Dewar, the dish itself, and the sub-reflector support legs in the signal path add their own polarization characteristics. All of these could, in theory, be included in the calibration described in this paper, however the generation of suitable calibration signals in the far-field of the antenna is very challenging. Astronomical calibration sources tend to be very weakly polarized. Further, the polarization characteristics of the antenna structure will undoubtedly change with gravitational deformation as a function of elevation angle, adding another dimension to the calibration problem. Therefore, in most practical cases which the authors can foresee, the raw performance of the digital orthomode transducer itself will simply serve to make its contribution to the overall polarization performance of the telescope negligible when compared to these other factors.

None of these issues are new or unique to the digital orthomode transducer design, as other types of orthomode transducers are also subject to the same limitations, but they are worth acknowledging whenever one claims performance with this level of precision.

## VIII. Design Constraints

The calibration procedure described in Section IV is equally applicable to other orthomode transducer formats than the planar probe design presented here. Any geometry that samples the EM field with at least two degrees of freedom should be sufficient. The key is that the orthomode transducer geometry provides minimal coupling to orthogonal or higher-order modes, and that what coupling is present is stable. The planar probe arrangement is convenient due to its compactness, the simplicity of manufacturing it, and the ability to connect the LNA directly to the terminal of the probe with minimal loss. Wider bandwidth, perhaps up to an octave, may be achievable with similar configurations, if a way can be found to suppress the higher order modes in the waveguide. So long as the probes couple well enough into the receiver chain to ensure good signal-to-noise ratio, the level of isolation and axial ratio over broader bandwidths should be the same as that reported here.

Theoretically, however, the LNAs could just as well be connected to the terminals of a differential, log-periodic antenna or the output of a conventional orthomode transducer like a quad-ridge or turnstile-junction-based design. The techniques described in this paper may be applied without modification to those designs as well for enhanced performance. Whatever the digital orthomode transducer geometry, stability of the analog electronic package is paramount to ensure the best possible precision between calibrations.

An attractive possibility for future designs is shown in Fig. 20a. Having only two probes in the circular waveguide minimizes the number of receiver channels and thus the required digital hardware – it would have the same digital bit rate as any conventional dual-polarized receiver for a given processed bandwidth. Such a design was not pursued with smooth-walled circular waveguide because the lack of rotational symmetry would permit mode-coupling between the $TE_{11}$ modes and $TM_{01}$ mode shown in Fig. 3. Although in theory this could be calibrated out, there was concern that external reflections of the $TM_{01}$ mode would not be stable. The design shown in Fig. 20a gets around this by modifying the waveguide geometry. The fluted walls push the cutoff frequency of the $TM_{01}$ mode above the standard waveguide band, while giving up nothing in compactness (unlike a quad-ridge orthomode transducer, which cuts off higher order modes at the expense of significant physical length [6]).

A second possibility for future enhancement is shown in Fig. 20b. In this modification, a concentric, coaxial probe is inserted into the waveguide through the backshort. This center-probe would not couple into any propagating mode in the waveguide, but it would leak a small amount of power into all the planar probes equally. This could be useful as a test port to inject noise for gain and system temperature calibrations, as well as for introducing CW tones for the purposes of monitoring aging effects in the receiver chain. That information in turn could be used to update the polarization and sideband calibrations in the field, if needed.

## IX. Conclusion

A novel type of planar orthomode transducer has been developed which takes advantage of numerical digital processing to reconstruct the incoming linear and circular polarizations without the aid of analog baluns or hybrids. A rigorous calibration algorithm was described which can accurately determine the amplitude and phase response of arbitrarily located probes, and automatically corrects for potential angular misalignments in the calibrators. Two prototype units were constructed, one with three probes in a circular waveguide, and another with four probes. With the aid of digital calibration, these units achieved better than 50 dB linear polarization isolation and better than 0.2 dB axial ratio for circular polarization, both stable over a temperature range of at least 10 C.

## Acknowledgment

The authors wish to thank their colleagues Anthony Kerr, Shing-Kuo Pan, Marian Pospieszalski, and John Webber for their valued advice and discussion. We also thank Francois Johnson for her assistance with IF board assembly.

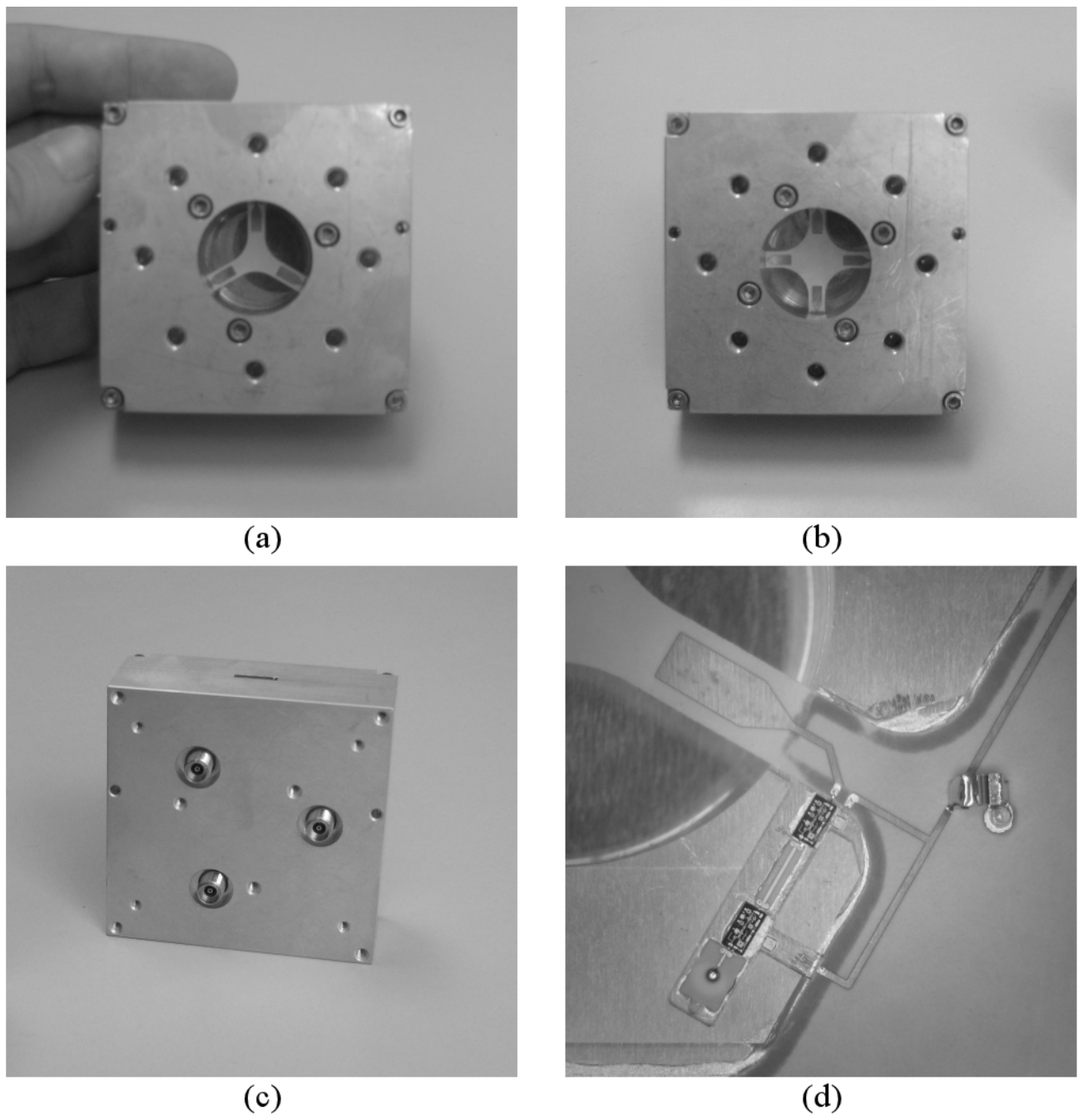

Fig. 1. Photographs of the prototype digital orthomode transducers. a) Three-probe digital orthomode transducer input side. b) Four-probe digital orthomode transducer input side. c) Three-probe digital orthomode transducer output side. d) Close-up of interior showing a single probe with attached MMIC amplifiers.

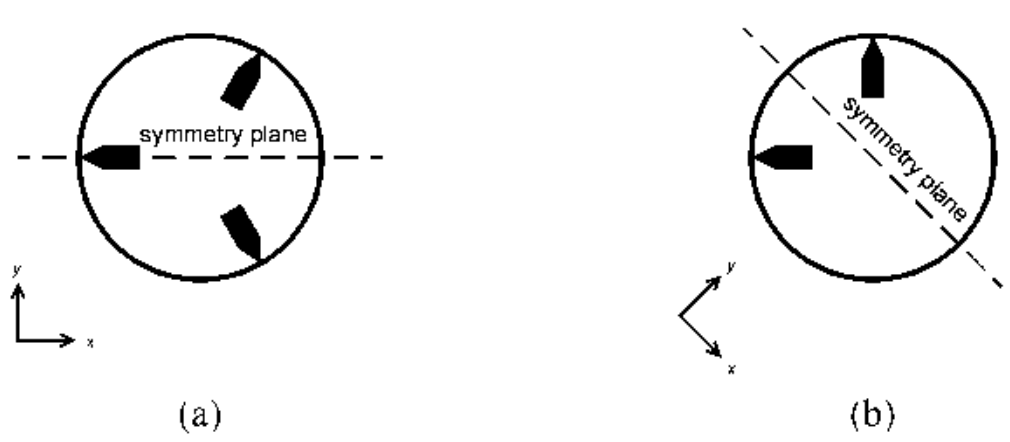

Fig. 2. Diagram of a) three-probe and b) two-probe digital orthomode transducer structure with the *x* axis defined along the symmetry plane.

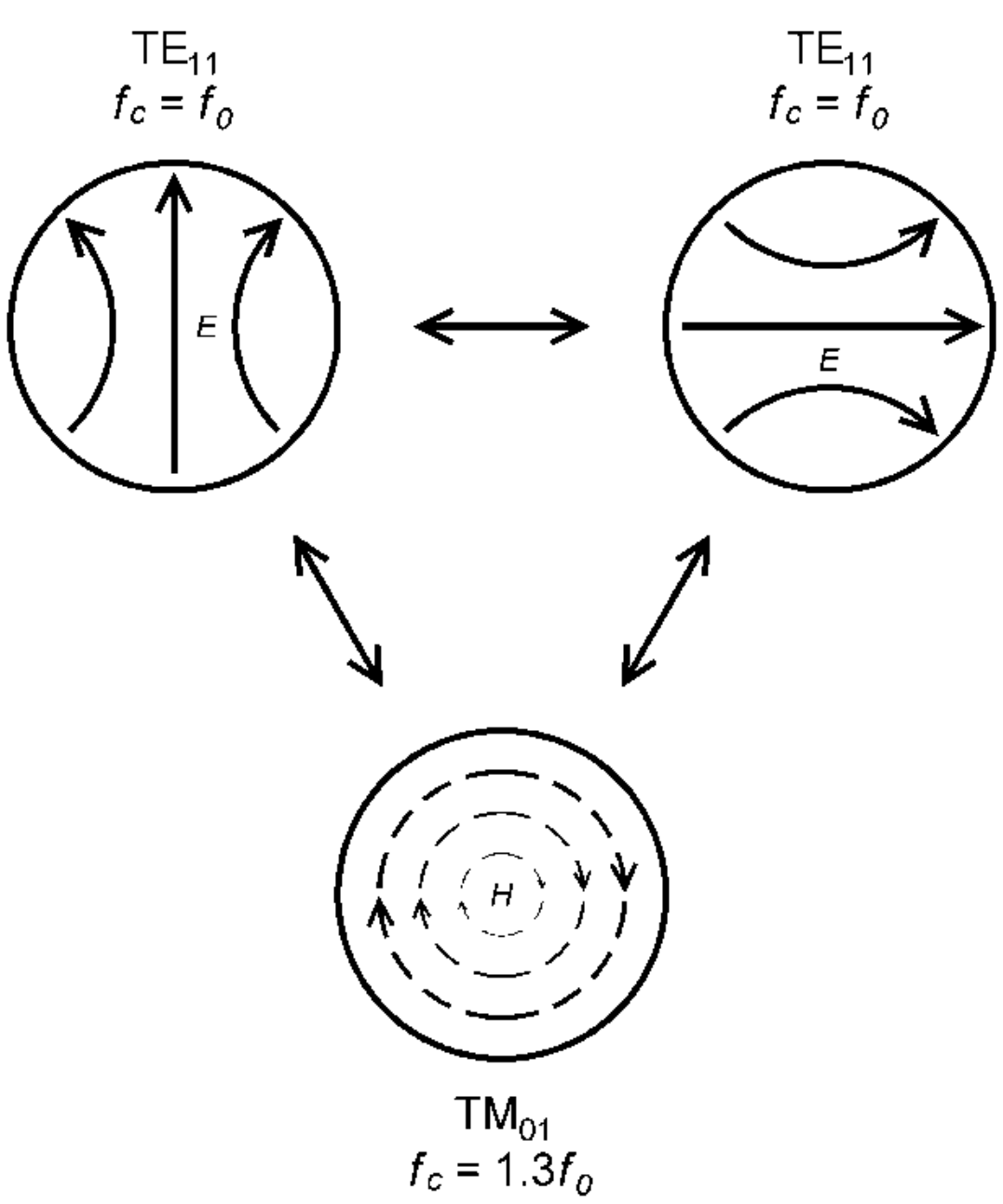

Fig. 3. Field diagrams of the first three modes in circular waveguide. $f_0$ is the cutoff frequency of the two dominant modes.

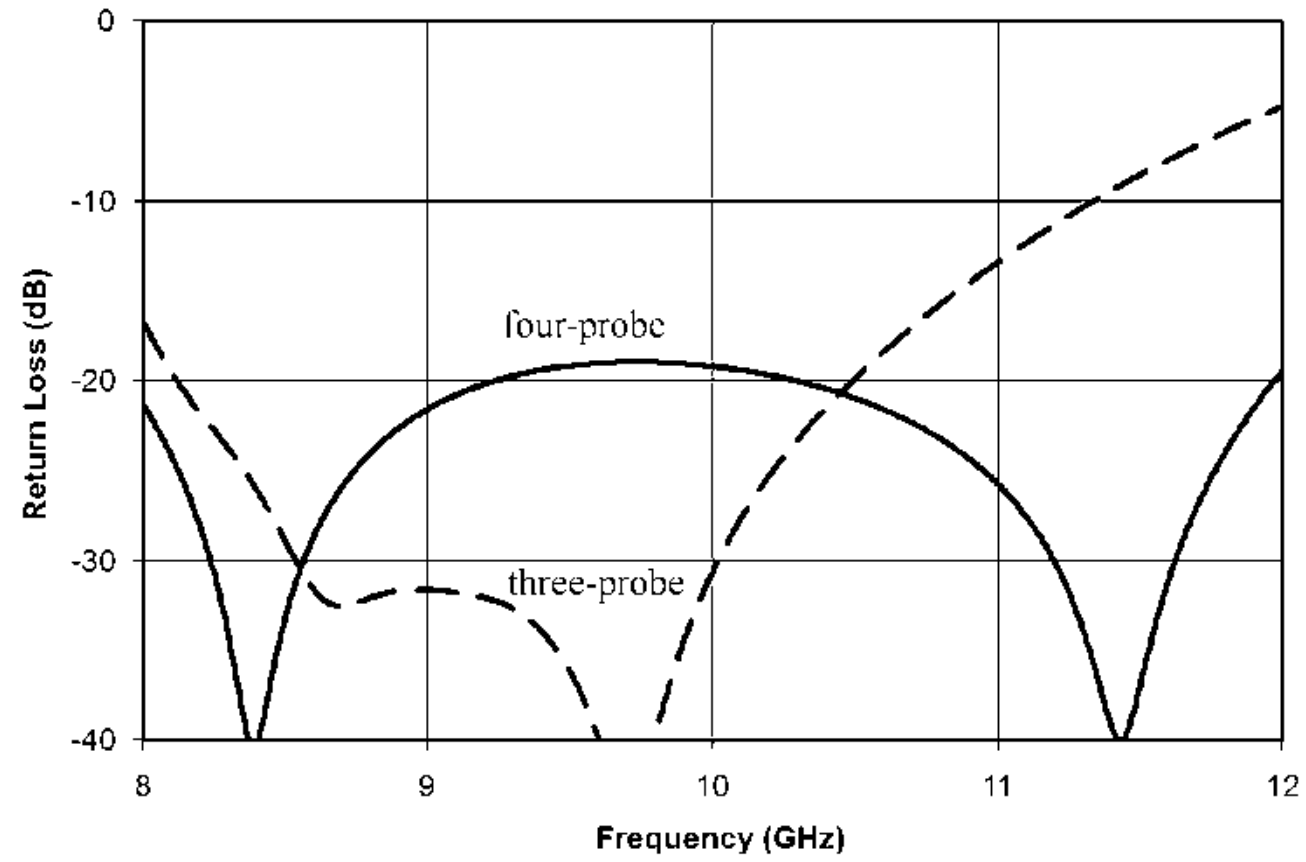

Fig. 4. Simulated return loss for the three-probe and four-probe digital orthomode transducers as seen from the circular waveguide port, assuming 50 Ω terminations on all probes. The actual return loss will be dominated by the input impedance of the LNAs.

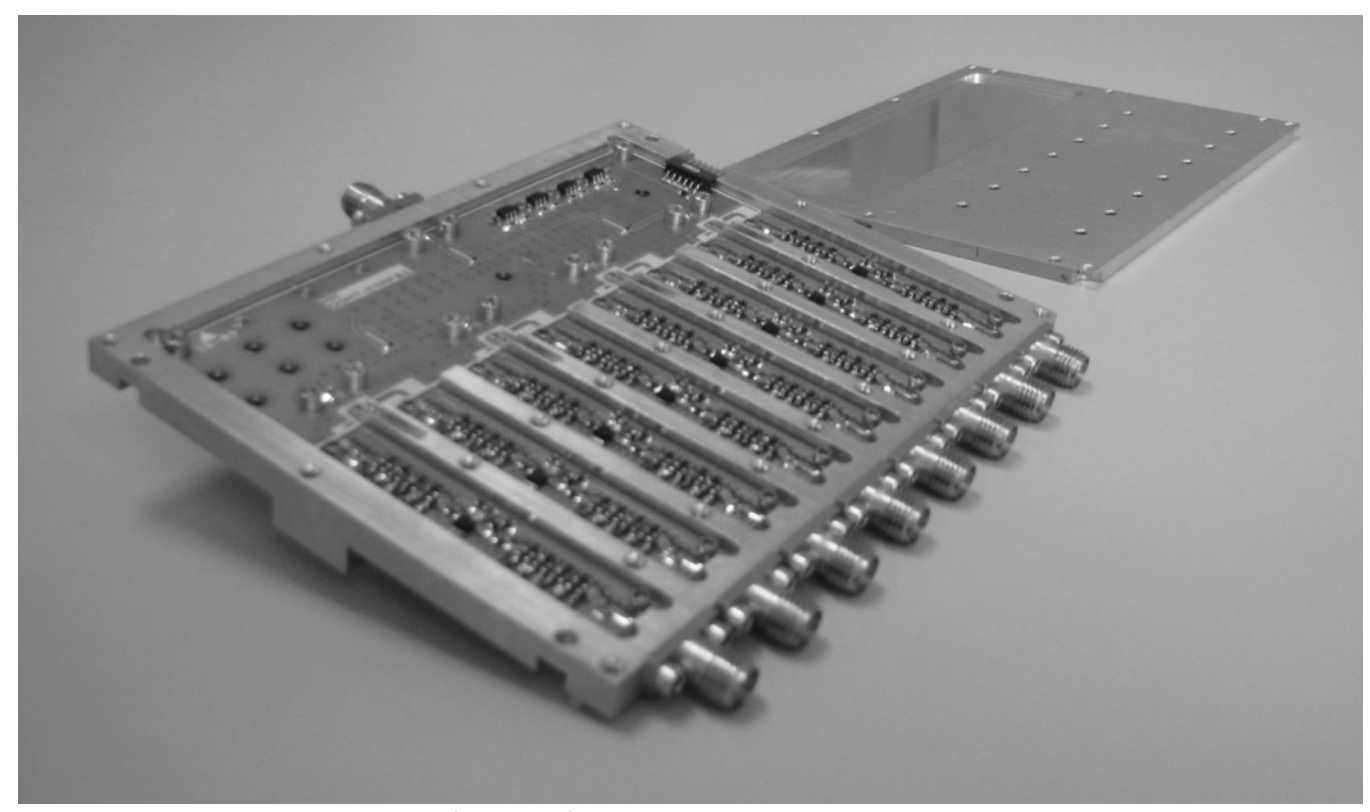

Fig. 5. Photograph of the four-channel downconverter unit employing digital sideband separating mixers (DSSMs).

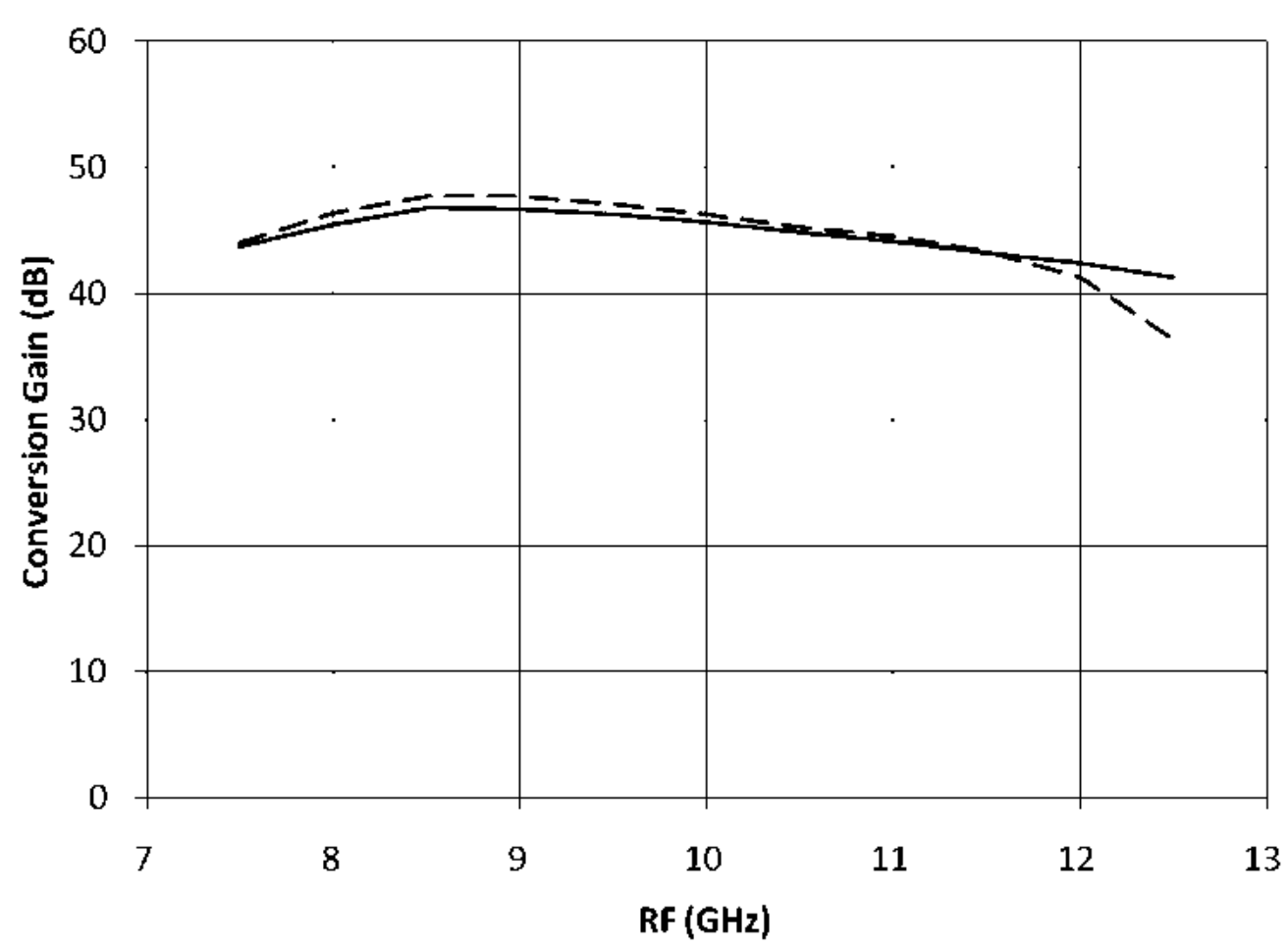

Fig. 6. Measured conversion gain versus RF for the downconverter, using the I and Q outputs for a typical channel. The IF was 10 MHz.

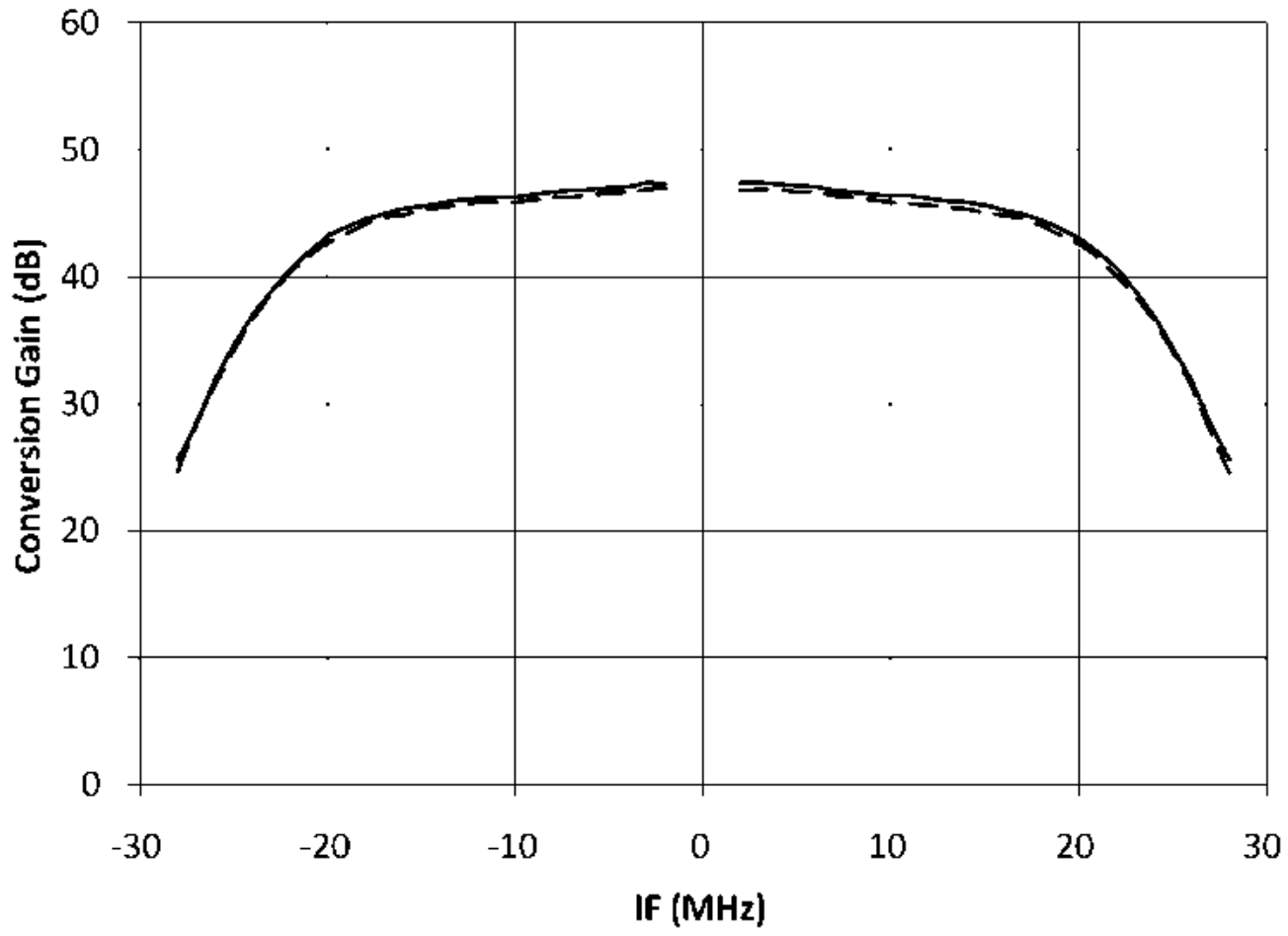

Fig. 7. Measured conversion gain versus IF for the downconverter, using the I and Q outputs for a typical channel. The LO was set to +13 dBm at 10 GHz.

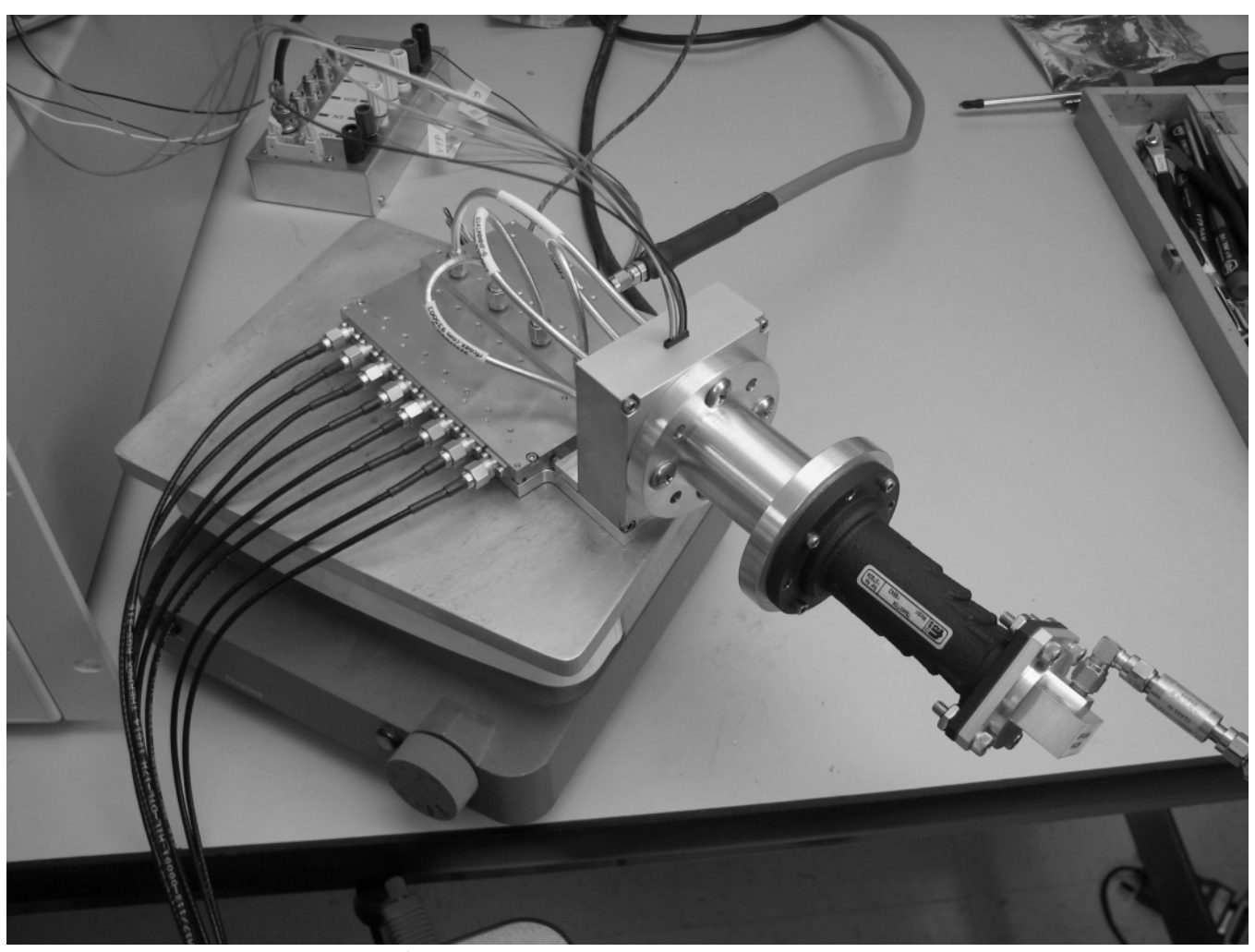

Fig. 8. Photograph of the digital orthomode transducer test setup. The entire test receiver was mounted on a hot plate to control the temperature and test the calibration stability.

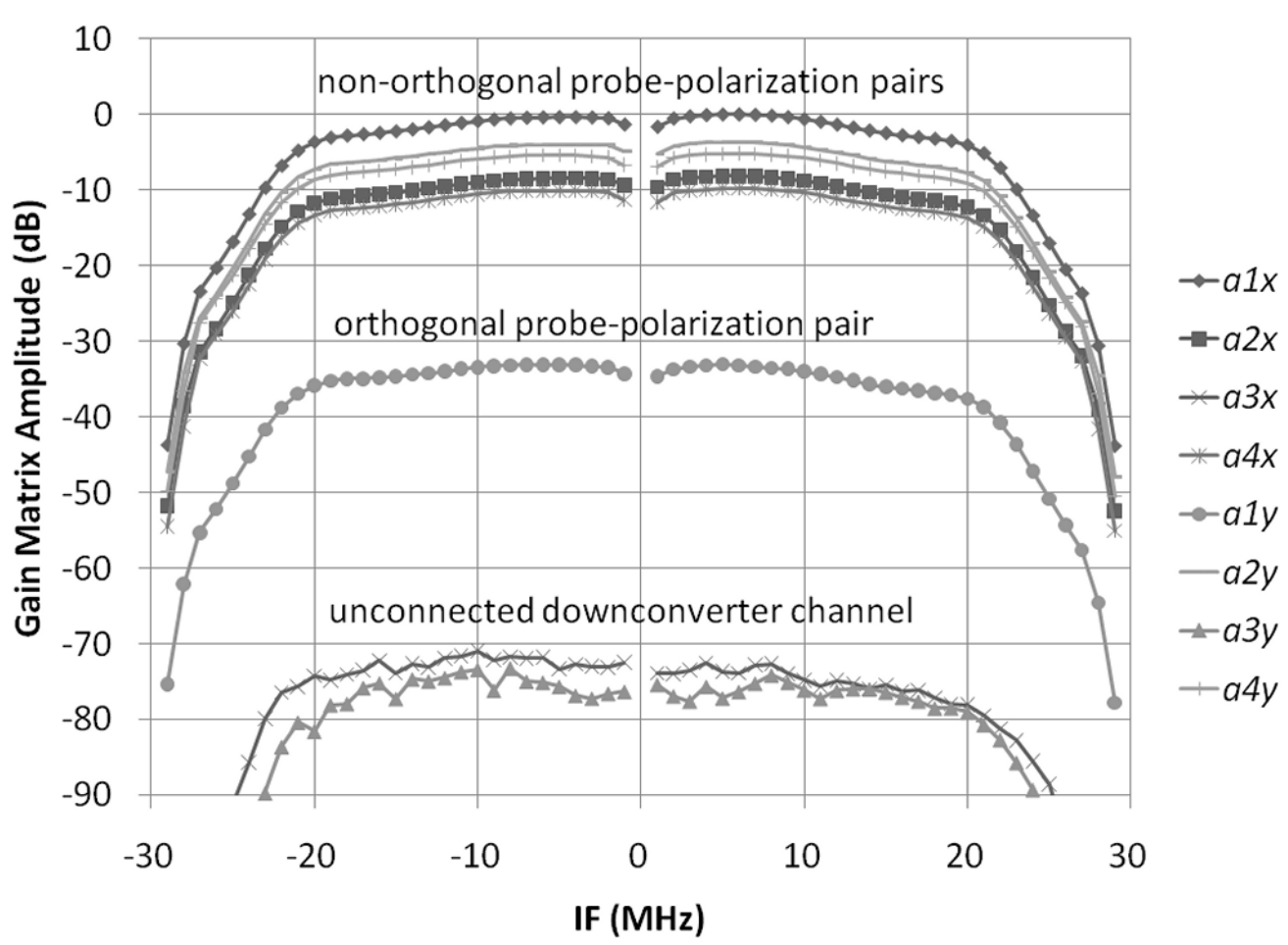

Fig. 9. Calibrated gain matrix amplitudes for three-probe digital orthomode transducer. Channel three on the four-channel downconverter was left unconnected, but picks up leakage from the other channels at the -70 dB level.

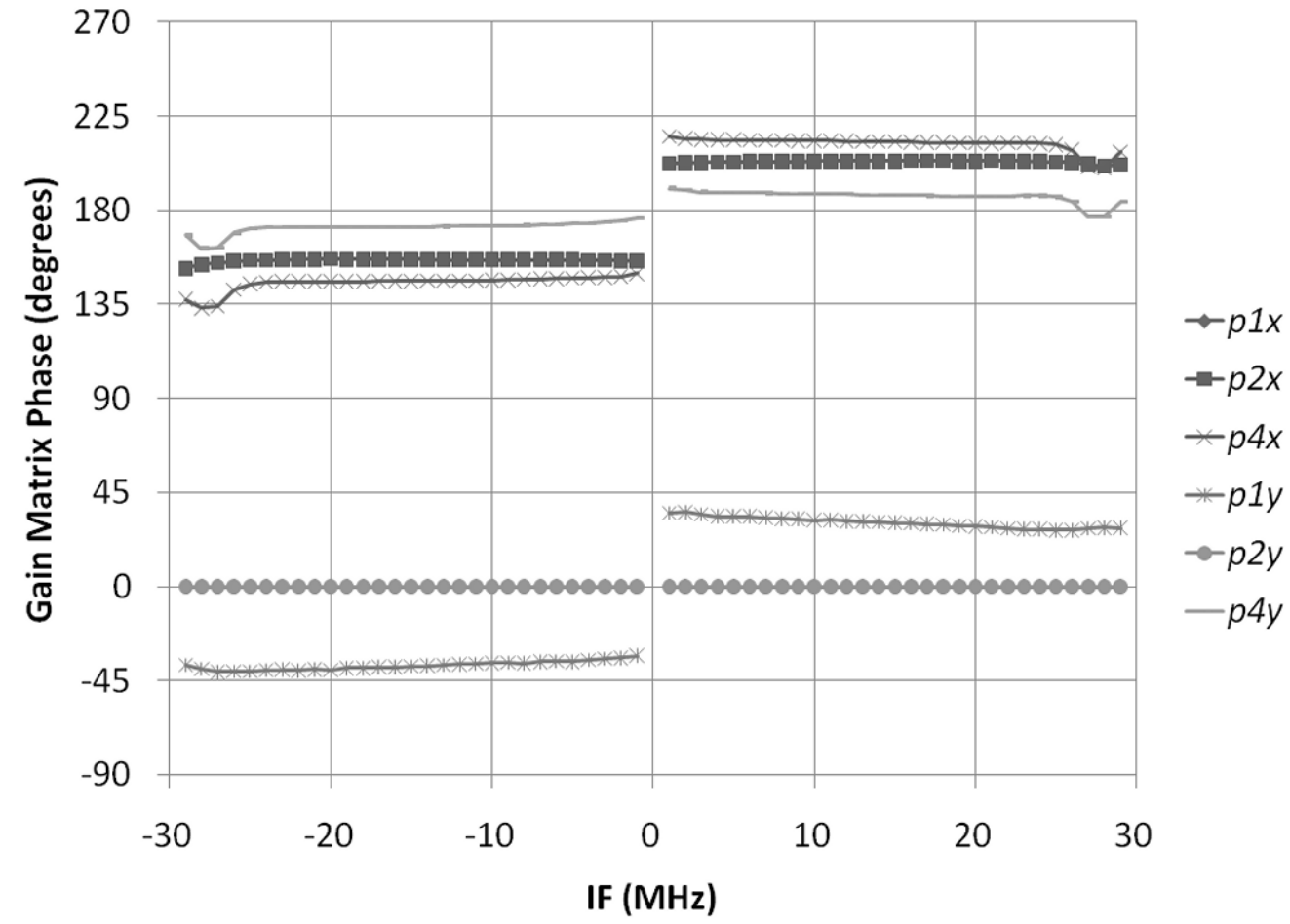

Fig. 10. Calibrated gain matrix phases for three-probe digital orthomode transducer.

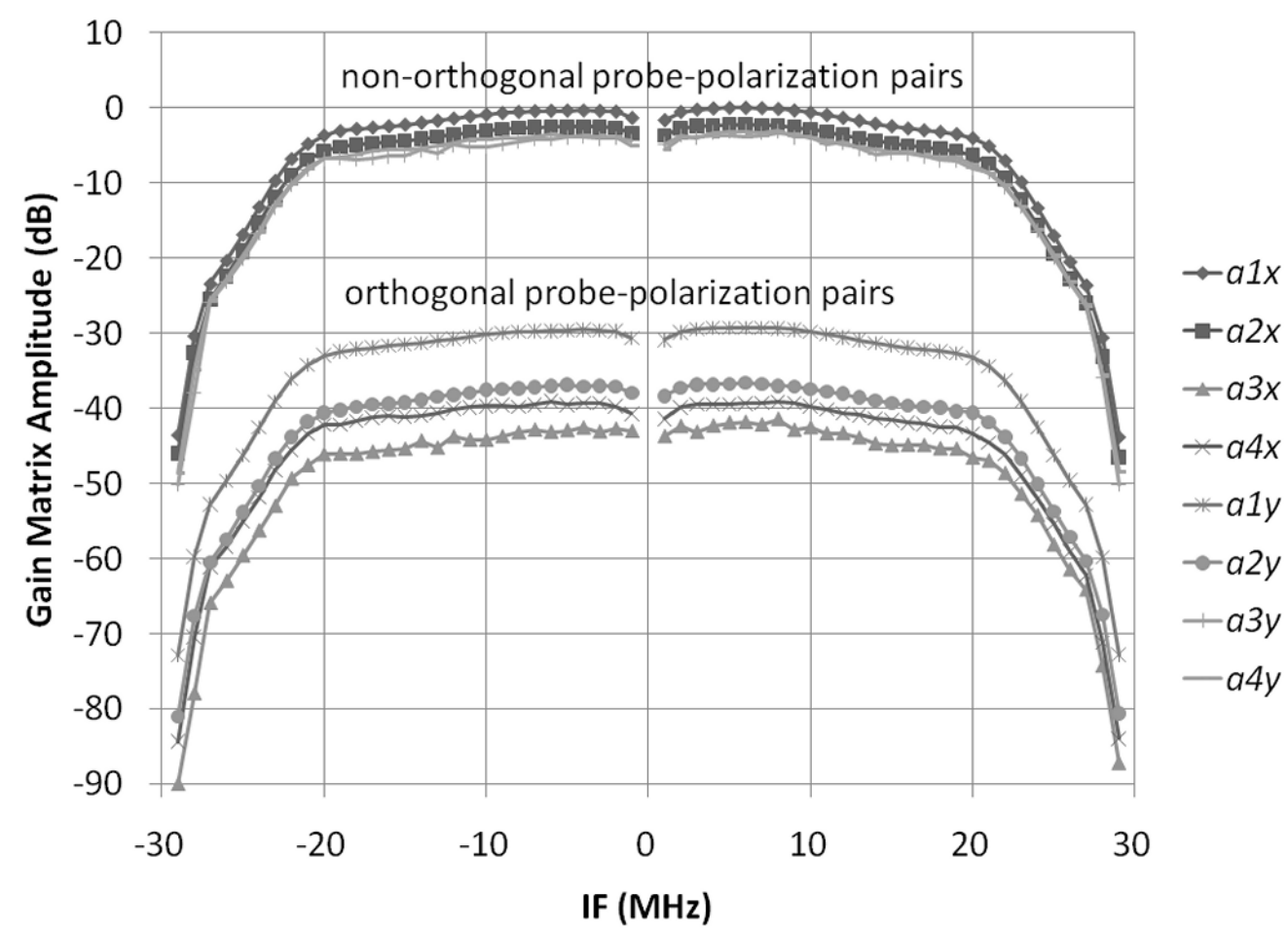

Fig. 11. Calibrated gain matrix amplitudes for four-probe digital orthomode transducer. All channels on the downconverter were used.

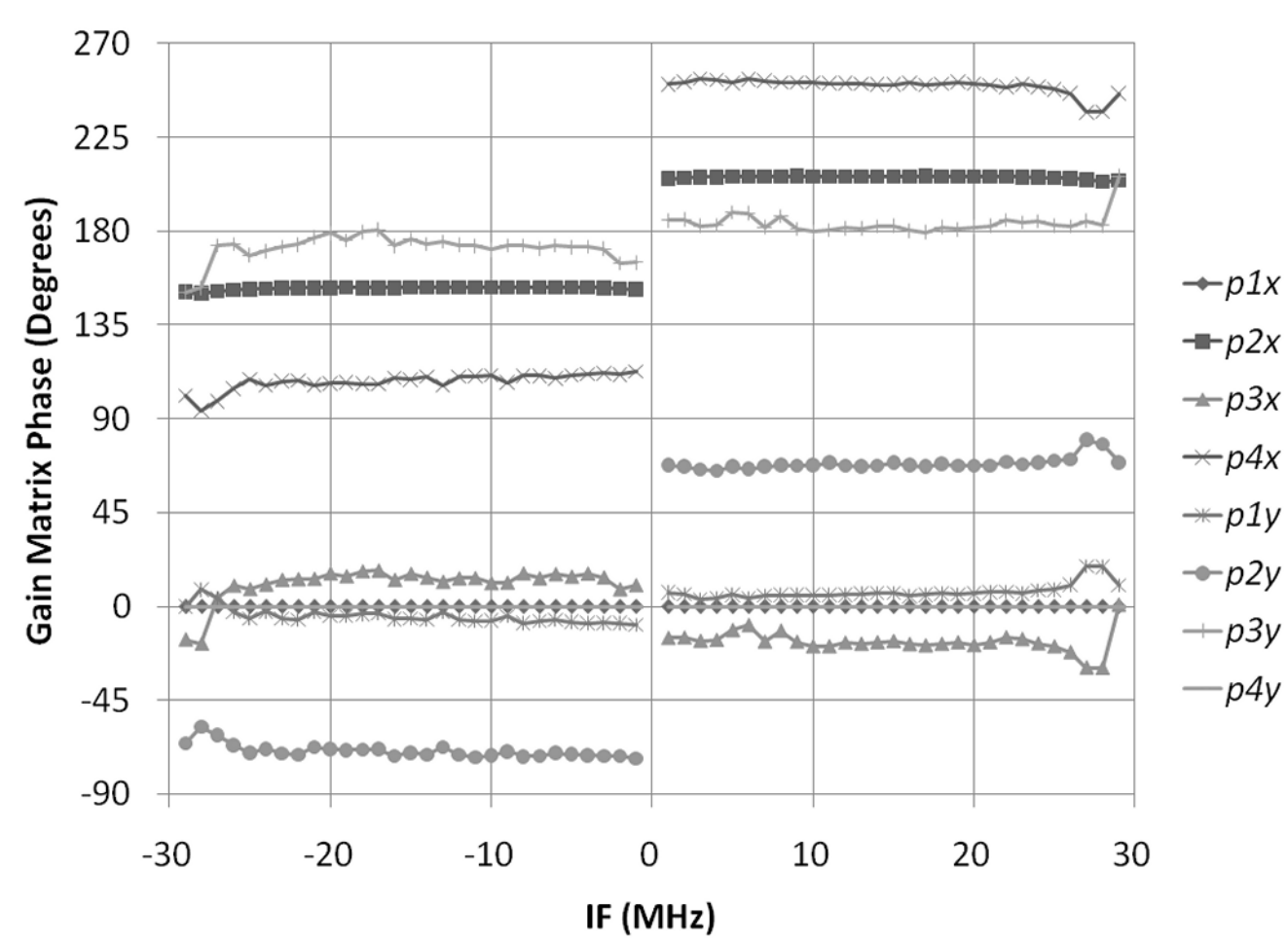

Fig. 12. Calibrated gain matrix phases for four-probe digital orthomode transducer.

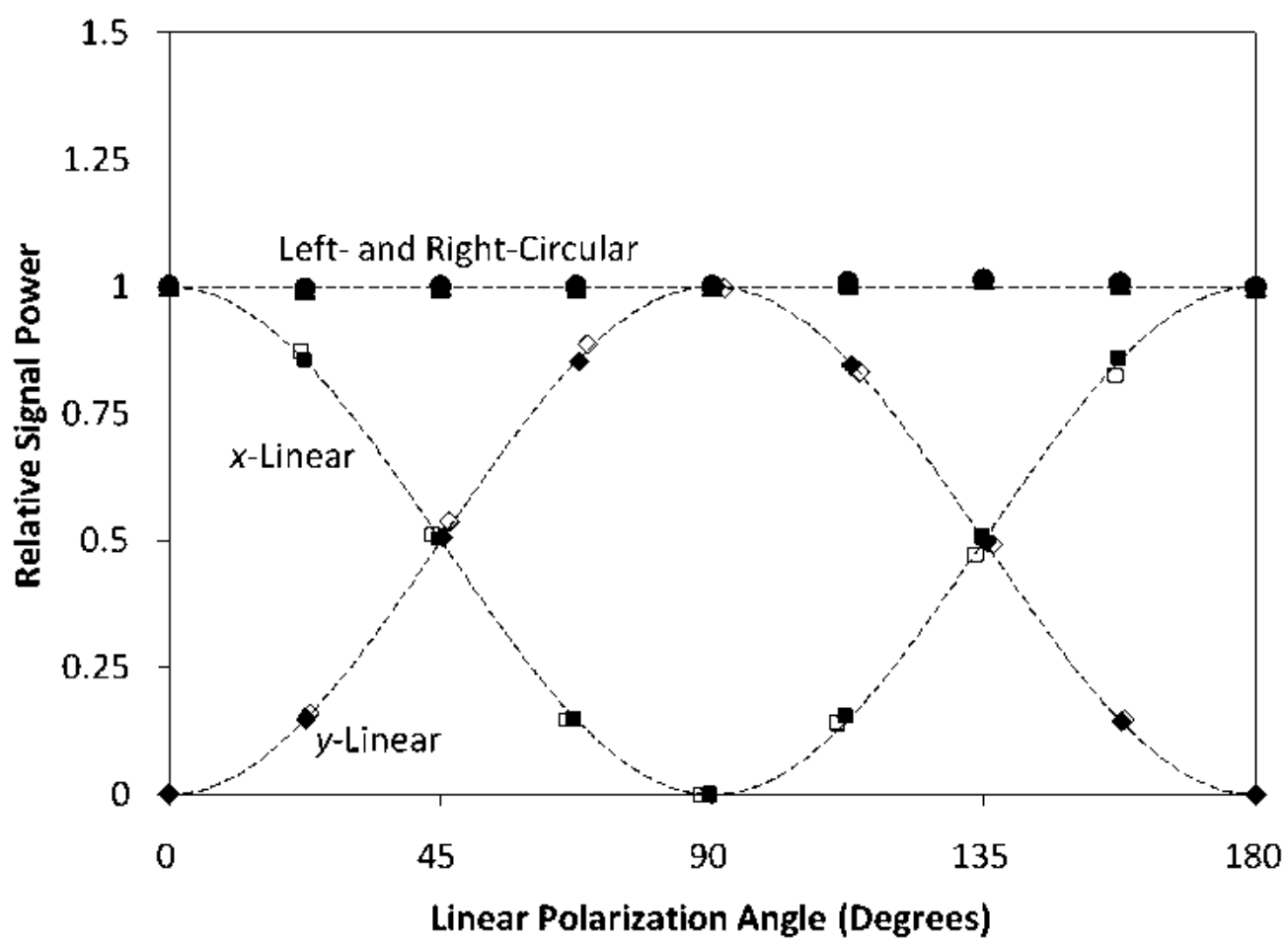

Fig. 13. Synthesized linear and circular polarizations as a function of the linear polarization angle set by the input transition. Markers show the measured data for both digital orthomode transducers, while the dashed lines show the theoretical prediction. Each point is averaged across IF frequency.

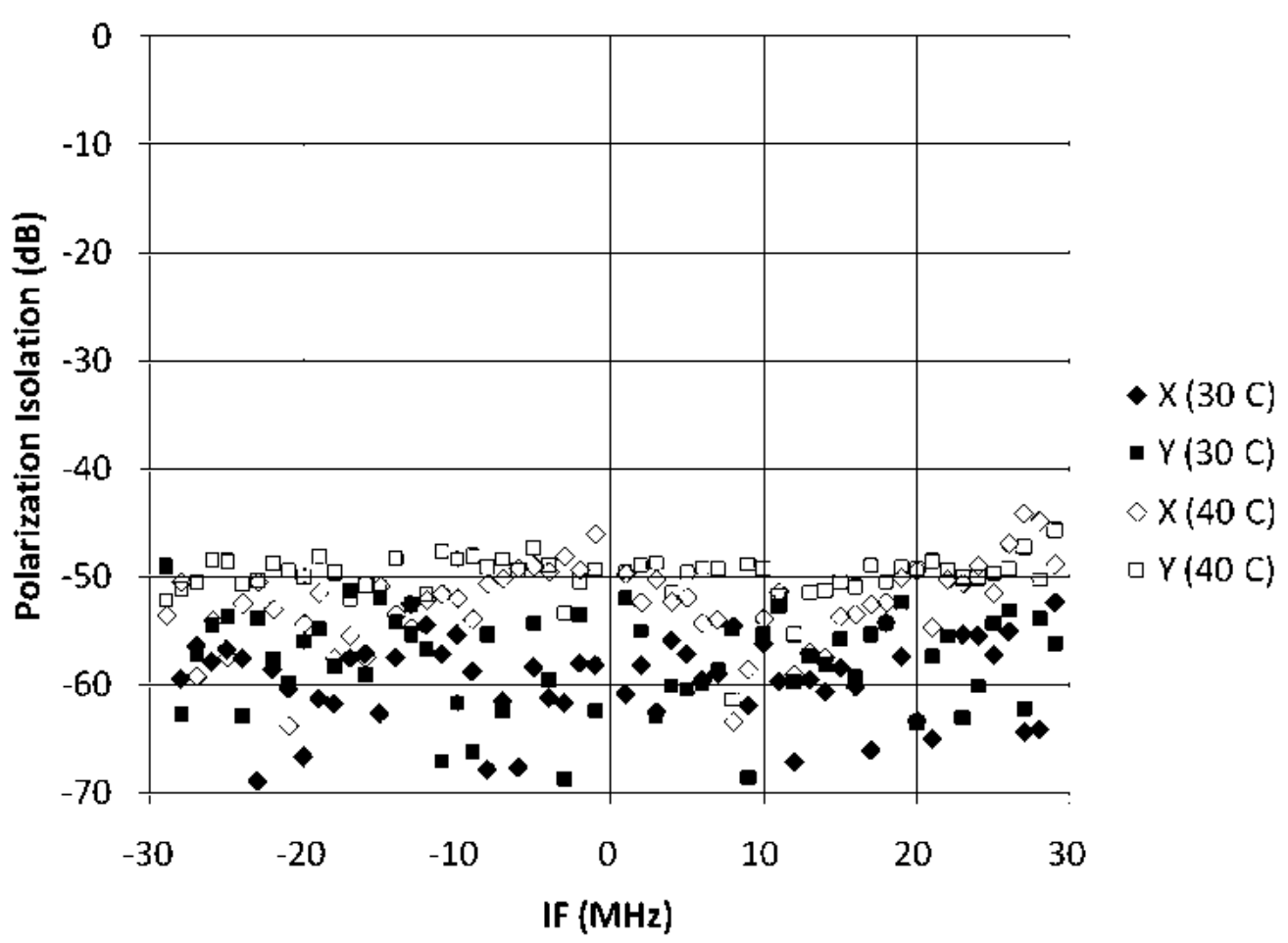

Fig. 14. Measured polarization isolation for the three-probe digital orthomode transducer. The measured result immediately after the calibration is shown with the closed markers. The measurement was then repeated after changing the base plate temperature by 10 C without recalibrating, and is shown with the open markers.

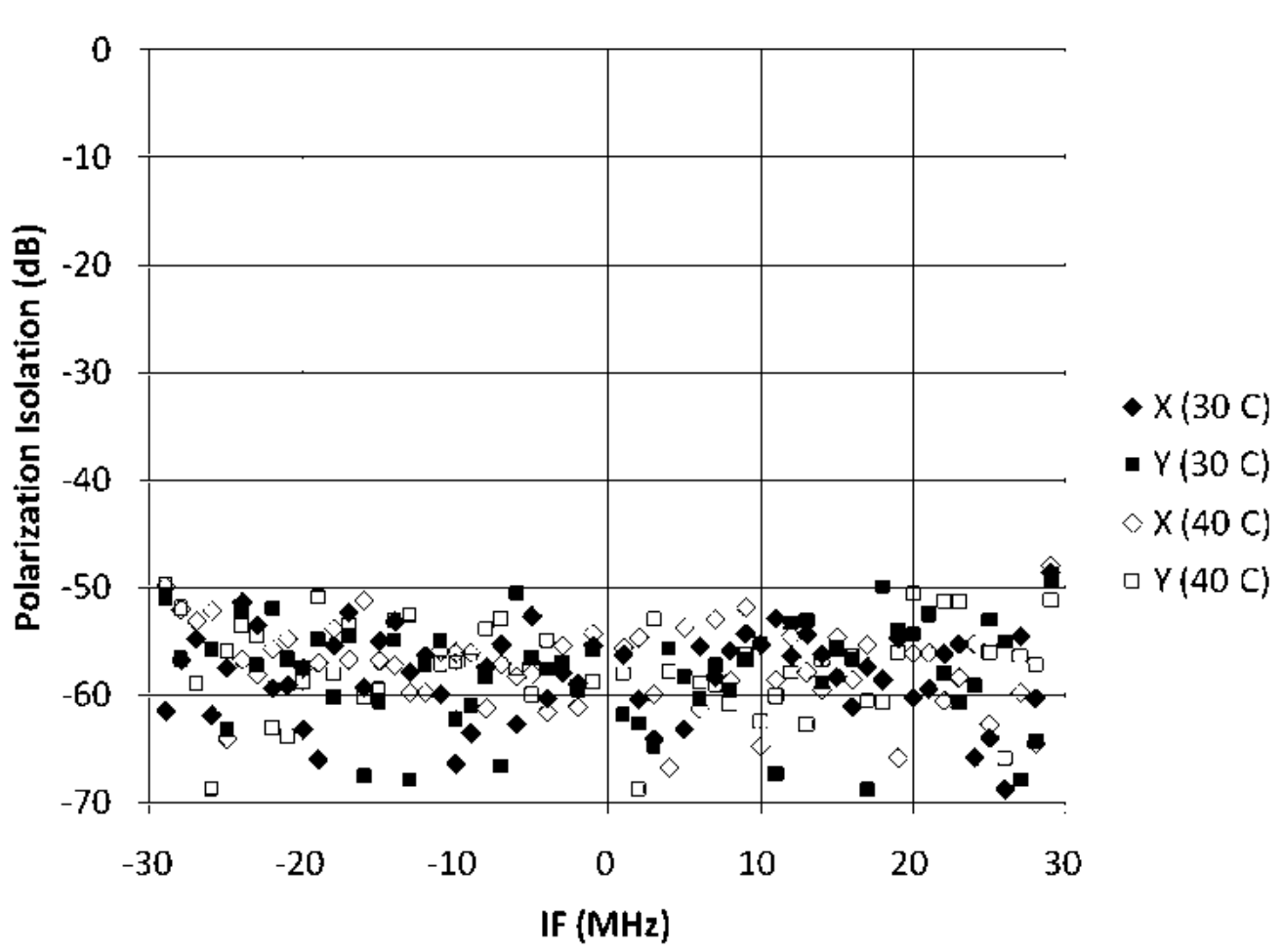

Fig. 15. Measured polarization isolation for the four-probe digital orthomode transducer. The measured result immediately after the calibration is shown with the closed markers. The measurement was then repeated after changing the base plate temperature by 10 C without recalibrating, and is shown with the open markers.

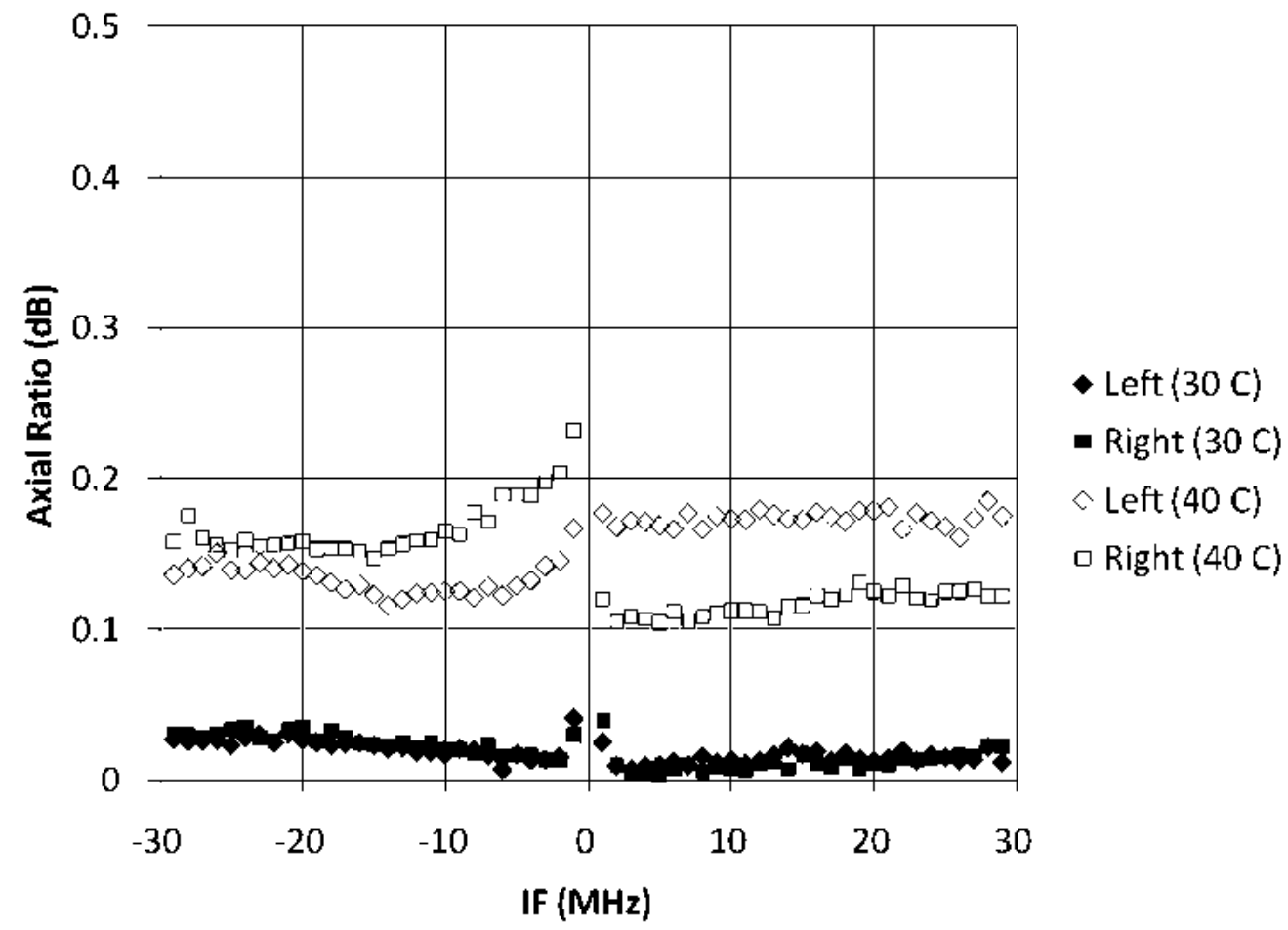

Fig. 16. Measured axial ratio for the three-probe digital orthomode transducer. The measured result immediately after the calibration is shown with the closed markers. The measurement was then repeated after changing the base plate temperature by 10 C without recalibrating, and is shown with the open markers.

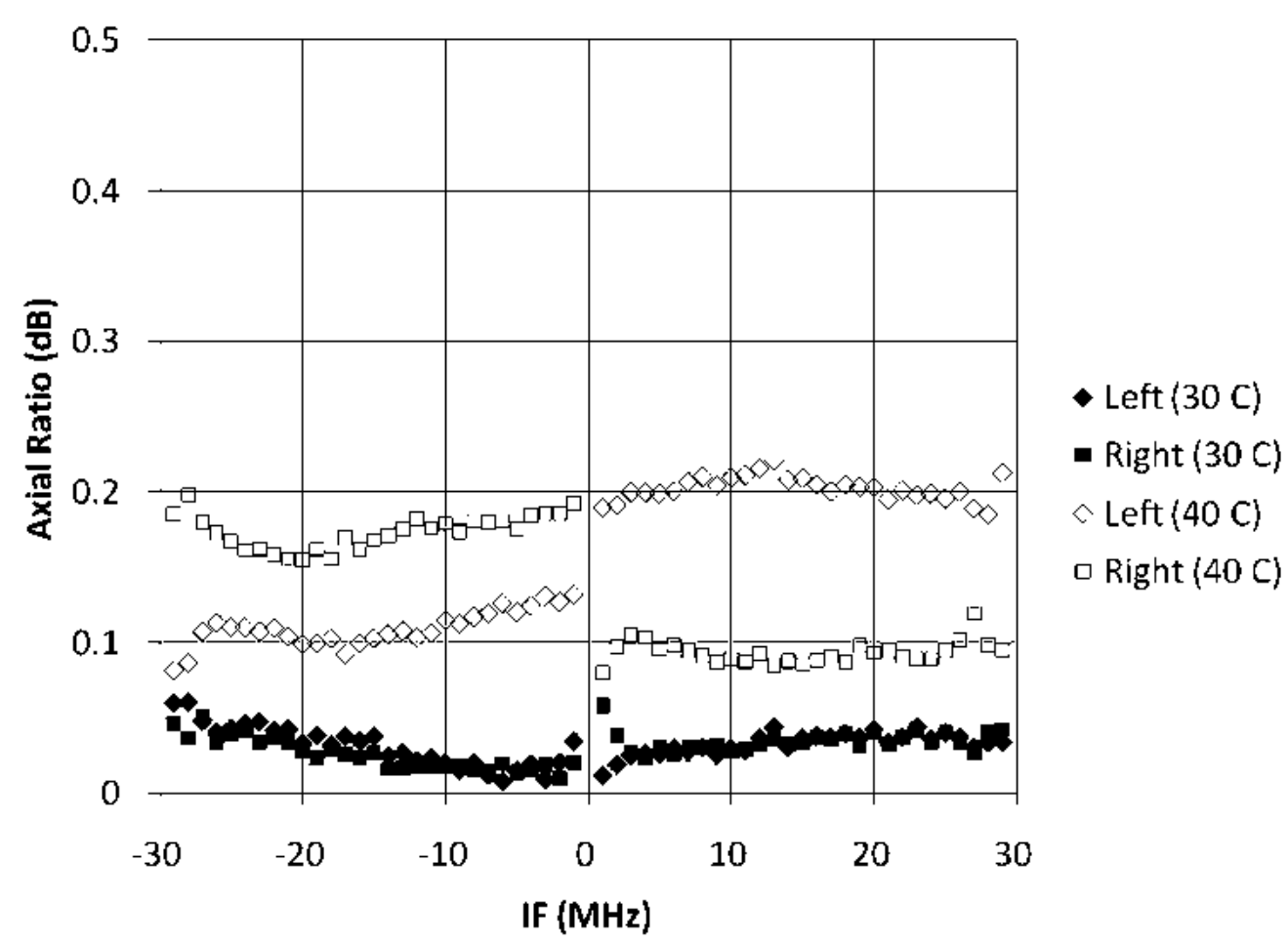

Fig. 17. Measured axial ratio for the four-probe digital orthomode transducer. The measured result immediately after the calibration is shown with the closed markers. The measurement was then repeated after changing the base plate temperature by 10 C without recalibrating, and is shown with the open markers.

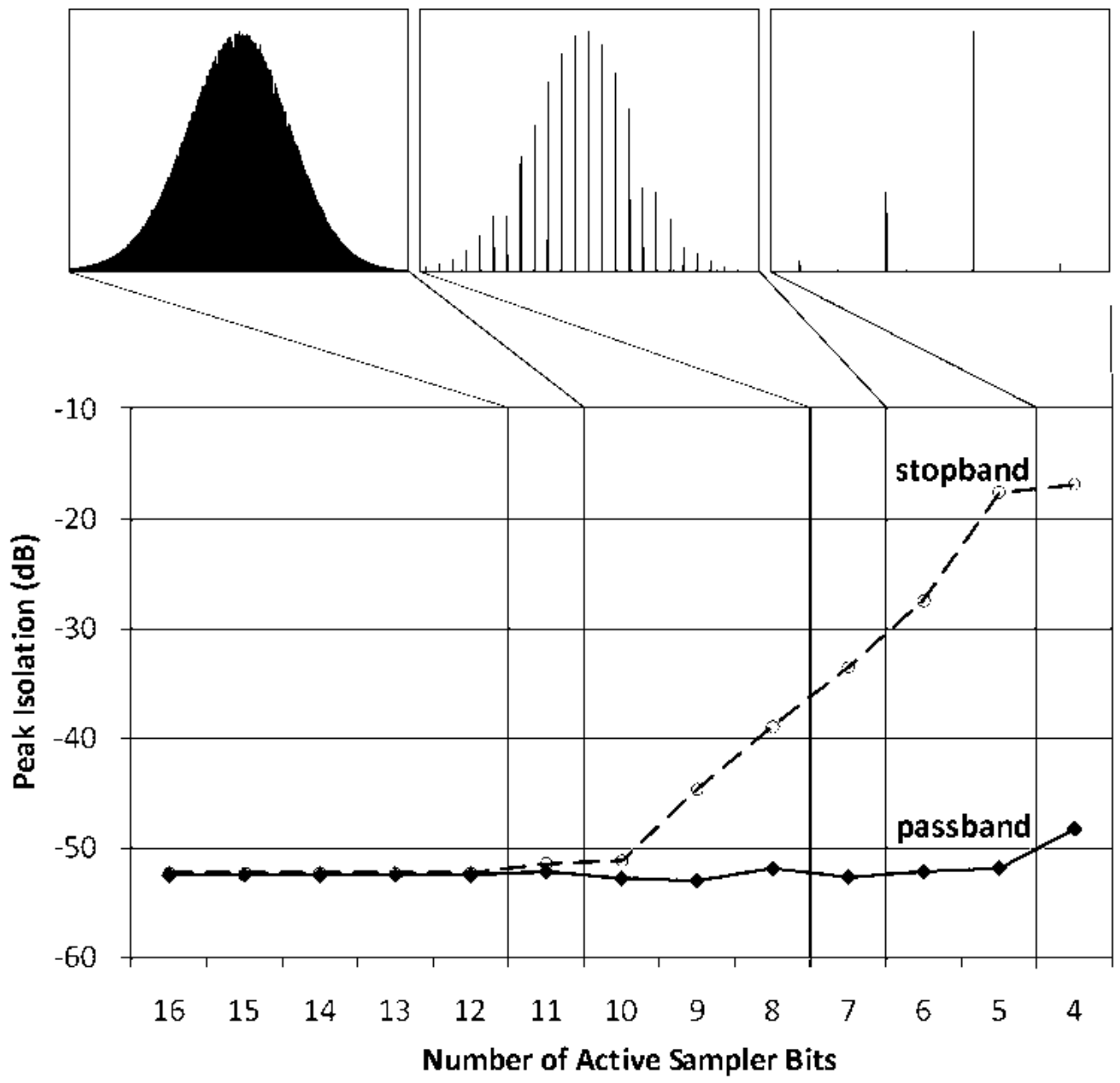

Fig. 18. Measured peak isolation in the passband (IF<24 MHz) and in the stopband (IF=24-29 MHz) for the three-probe digital orthomode transducer. The abscissa shows the number of unmasked bits. The insets above the plot show histograms of the data values after masking for select bit resolutions. The analog signal level was such that with only 4 active bits the ADC behaved like a 1 bit (2 level) sampler.

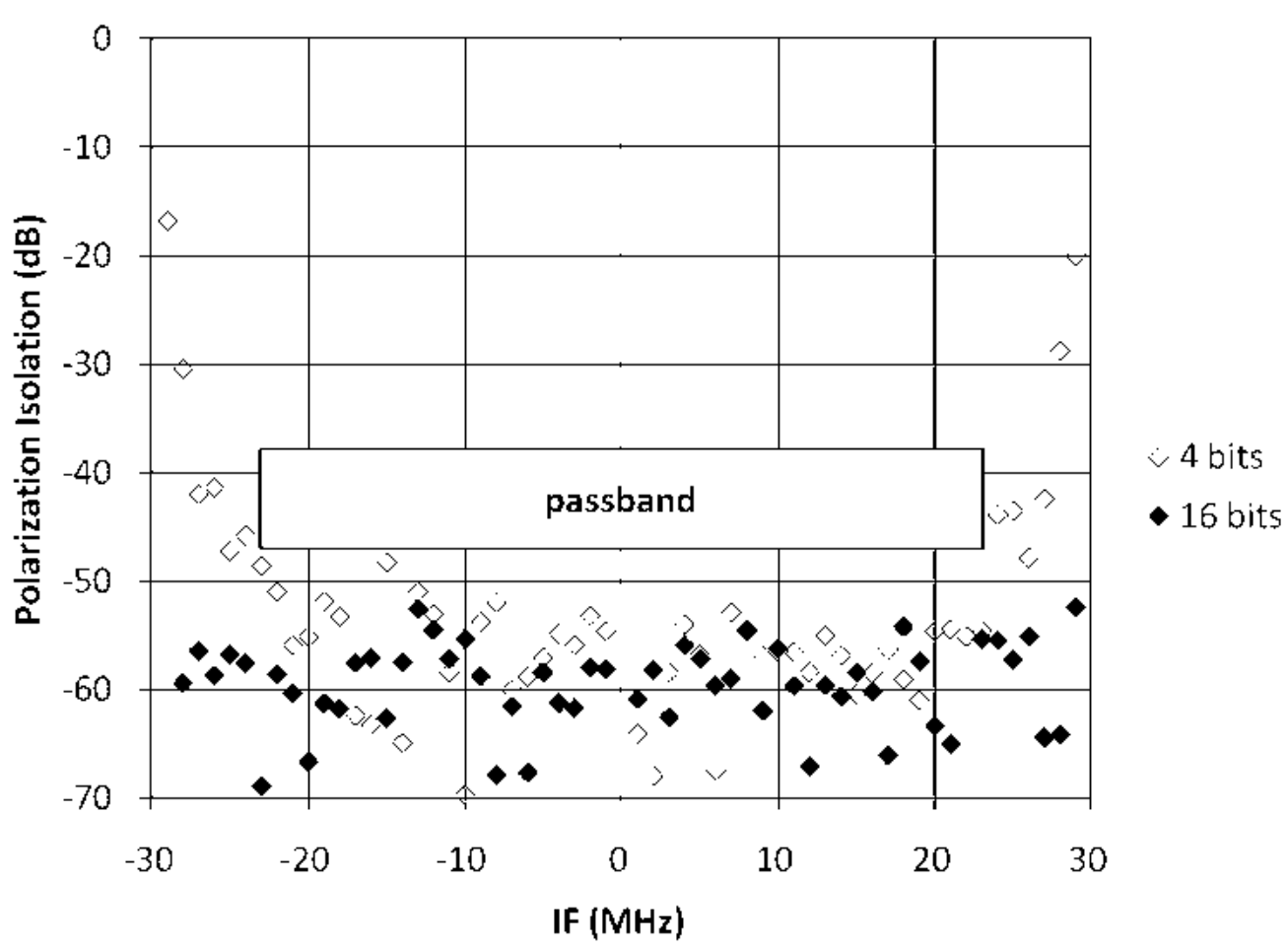

Fig. 19. Measured polarization isolation for the three-probe digital orthomode transducer with full 16 bit resolution (closed markers) and with all but 4 bits masked off (open markers). The analog signal level was such that with all but 4 bits masked off, the ADC behaved like a 1 bit (2 level) sampler.

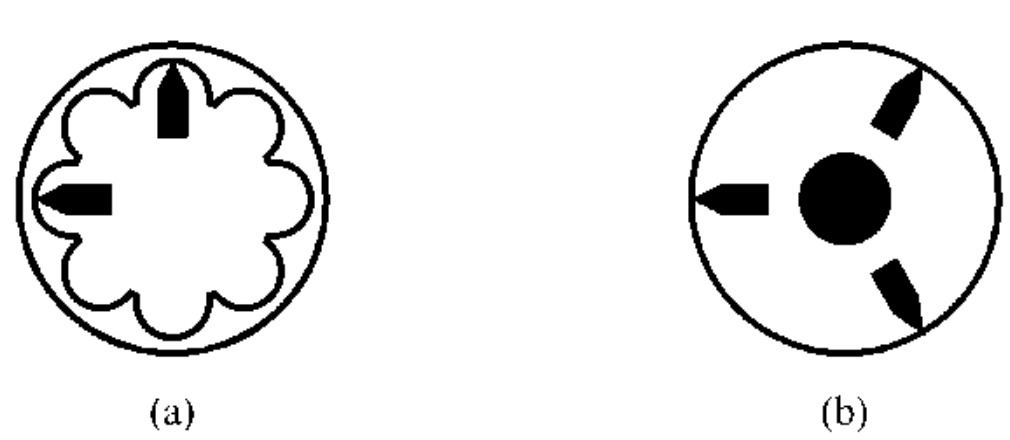

Fig. 20. Diagram of two possible future enhancements. a) Two-probe digital orthomode transducer with fluted waveguide to suppress the $TM_{01}$ mode. b) Three-probe digital orthomode transducer with coaxial center-probe for calibration injection.